\documentclass[12pt]{article}
\usepackage{titling}

\usepackage[left=20mm, right=20mm, top=15mm, bottom=20mm, columnsep=5mm]{geometry}

\usepackage[english]{babel}
\usepackage[utf8]{inputenc}
\usepackage[T1]{fontenc}
\usepackage{xcolor}

\usepackage{tabularx}
\usepackage{enumerate}
\usepackage{multirow}
\usepackage{hyperref}
\usepackage{booktabs}

\usepackage{amsmath}
\usepackage{amssymb}
\usepackage{amsfonts}
\DeclareMathSymbol{\shortminus}{\mathbin}{AMSa}{"39}

\usepackage{authblk}

\usepackage{graphicx}
\usepackage{float}

\usepackage[backend=bibtex,style=phys]{biblatex}
\bibliography{THI21}

\title{Asymmetric Adaptivity induces Recurrent Synchronization in Complex Networks }
\author[1]{Max Thiele}
\author[1,2,3]{Rico Berner}
\author[4]{Peter A. Tass}
\author[1,5,6]{Eckehard Schöll}
\author[6,2,7]{Serhiy Yanchuk}
\affil[1]{Institut für Theoretische Physik, Technische Universität Berlin, 10623 Berlin, Germany}
\affil[2]{Institut für Mathematik, Technische Universität Berlin, 10623 Berlin, Germany}
\affil[3]{Institut für Physik, Humboldt Universität zu Berlin, 10117 Berlin, Germany}
\affil[4]{Stanford University, Department of Neurosurgery, Stanford, CA, 94305, USA}
\affil[5]{Bernstein Center for Computational Neuroscience Berlin, 10115 Berlin, Germany}
\affil[6]{Potsdam Institute for Climate Impact Research, 14473 Potsdam, Germany}
\affil[7]{Institut für Mathematik, Humboldt Universität zu Berlin, 10117 Berlin, Germany}

\date{}

\begin{document}

\maketitle

\begin{abstract}
    Rhythmic activities that alternate between coherent and incoherent phases are ubiquitous in chemical, ecological, climate, or neural systems. Despite their importance, general mechanisms for their emergence are little understood. In order to fill this gap, 
    we present a framework for describing the emergence of recurrent synchronization in complex networks with adaptive interactions. This phenomenon is manifested at the macroscopic level by temporal episodes of coherent and incoherent dynamics that alternate recurrently. At the same time, the dynamics of the individual nodes do not change qualitatively. We identify asymmetric adaptation rules and temporal separation between the adaptation and the dynamics of individual nodes as key features for the emergence of recurrent synchronization. Our results suggest that asymmetric adaptation might be a fundamental ingredient for recurrent synchronization phenomena as seen in pattern generators, e.g., in neuronal systems.
\end{abstract}

\section{Introduction}

A multitude of real-world dynamical networks possess adaptive interactions \cite{GRO06b,GRO08a,JAI01,GUT11,SAY13,BUS08}. 
Structural changes in such networks depend on the state of individual nodes. The state of the nodes in turn depends on the connectivity. 
This results in a feedback loop between the dynamics (function) and the network structure. In neuronal networks, for example, spike-timing-dependent plasticity, a special type of network adaptation \cite{BLI93,MAR97a,ABB00,BI01,CAP08a,CAT08,CLO10}, contributes to learning~\cite{BRO88a} or anti-kindling~\cite{TAS06} effects. 
Another adaptation mechanism, structural plasticity, is responsible for a homeostatic regulation of electrical activity in the brain~\cite{BUT13c}. Beyond neural networks, adaptive networks are used in communication systems~\cite{GAV12} and for modeling complex behavior in social, climate or ecological systems~\cite{MUE17,SAY13,NUW17}. 

In realistic systems, the adaptation rules are likely to be heterogeneous. 
For example, the adaptation rule in neuronal networks may depend on the distance between interacting neurons~\cite{FRO05a,SJO06}.
So far, little is known about the dynamical effects induced by heterogeneous adaptation~\cite{KAS21a,BER21f}. 
The question arises as to which extent heterogeneous adaptation produces new functionality and thus represents an important element for system behavior. 
In this work, we answer this question by showing that heterogeneous adaptation can be a determining ingredient for producing new collective network function. In particular, we describe how heterogeneous adaptation induces the phenomenon of~\emph{recurrent synchronization}. 

Recurrent synchronization, which we study here, is a macroscopic phenomenon in dynamical networks involving the recurrent switching between synchronous behavior, represented e.g. by phase-locking, and asynchronous behavior, represented e.g. by frequency clustering. During recurrent synchronization, a macroscopic observable exhibits bursting behavior, whereas the individual nodes are not required to burst at the microscopic level. In our study we consider the nodes to have simple oscillatory dynamics. Therefore recurrent synchronization as a macroscopic effect contrasts bursting found in neuronal networks \cite{BEL11a,TAS12a,BEL08,GER14a}, where single neurons can have alternating periods of quiescence and fast spiking.


Other studies \cite{BLA99b,STO02, PAN12, SCH08, GAS20} have reported the occurrence of a similar phenomenon, called collective bursting, in neuronal networks. An important difference of \cite{BLA99b,STO02, PAN12, SCH08, GAS20} from our work, however, is that in these studies collective bursting phenomenon is induced and observable on the microscopic level of individual neurons, whereas recurrent synchronization is not manifested by bursting on the microscopic level. Another adaptation-induced switching between phase-locking and periodic oscillation has been observed in~\cite{CIS20}, which is related to fold-homoclinic bursting~\cite{IZH00}. In a two-population excitatory-inhibitory network of quadratic integrate and fire neurons, a dynamical phenomenon was observed showing bursts of high frequency and high amplitude activity at a slow burst rate~\cite{BYR22}. In contrast to the recurrent synchronization considered here, the populations change their individual dynamics from oscillatory to steady state during the episodes of high activity and quiescence. This phenomenon of activity bursting has been also reported in a population of excitable units adaptively coupled to a pool of resources~\cite{FRA22}.

In a broader sense, the mechanism for recurrent synchronization, which we report here, is based on a recurrent slow dynamics of hidden variables that we relate to adaptive coupling weights between dynamical populations in this work. When adaptation is heterogeneous (asymmetric), the hidden variables lead to re-emergence of episodes of synchronization and desynchronization. Effects such as converse symmetry breaking~\cite{NIS16,MOL20} or asymmetry induced synchronization~\cite{MOL21} have been described only very recently and are about to change our understanding of "good synchronization conditions" for dynamical systems. In this light, recurrent synchronization adds a new layer of dynamical complexity that can be induced by asymmetry, and hence heterogeneity, in complex dynamical networks.


In our study, we employ models of different hierarchical organization. With this, we show that recurrent synchronization may emerge in dynamical models of different complexity. A fairly complex system consists of two populations of Hodgkin-Huxley neurons with different adaptation rules inside and between the populations.
The adaptation rules are given by spike-timing-dependent plasticity with different adaptation functions. 
We also propose reduced phenomenological models of two coupled Hodgkin-Huxley-neurons as well as phase oscillators equipped with slowly evolving coupling weights. We show that the reduced models are capable of describing the main dynamical features arising in the intra- and inter-population dynamics. While the more complex model is studied numerically, the reduced phenomenological models are analyzed in more detail by methods from geometric singular perturbation, averaging, and bifurcation theories. In this work, we propose a general methodology to study recurrent synchronization in systems with arbitrary complex dynamical nodes and continuous as well as noncontinuous adaptation rules.

Our study reveals collective mechanisms behind the appearance of recurrent synchronization. In particular, we explain the importance of the following ingredients for the emergence of recurrent synchronization: slow adaptation, i.e., the timescale separation between the adaptation and the individual neuronal dynamics; asymmetry of adaptation rules; recurrent (periodic/spiking) dynamics of the individual neurons. We hypothesize that asymmetric adaptivity might play a fundamental role in emergence and impairment of neuronal pattern generators.

\section{Results}\label{sec:results}
\subsection{Recurrent synchronization}\label{sec:collBurst}

In this section, we introduce the phenomenon of recurrent synchronization, where certain macroscopic observables display bursting behavior while the individual nodes emit single periodic
spikes. We show that such a behavior is caused by alternating episodes of synchronization and desynchronization in a dynamical network. 

\begin{figure}
    \centering
    \includegraphics[width=0.7\textwidth]{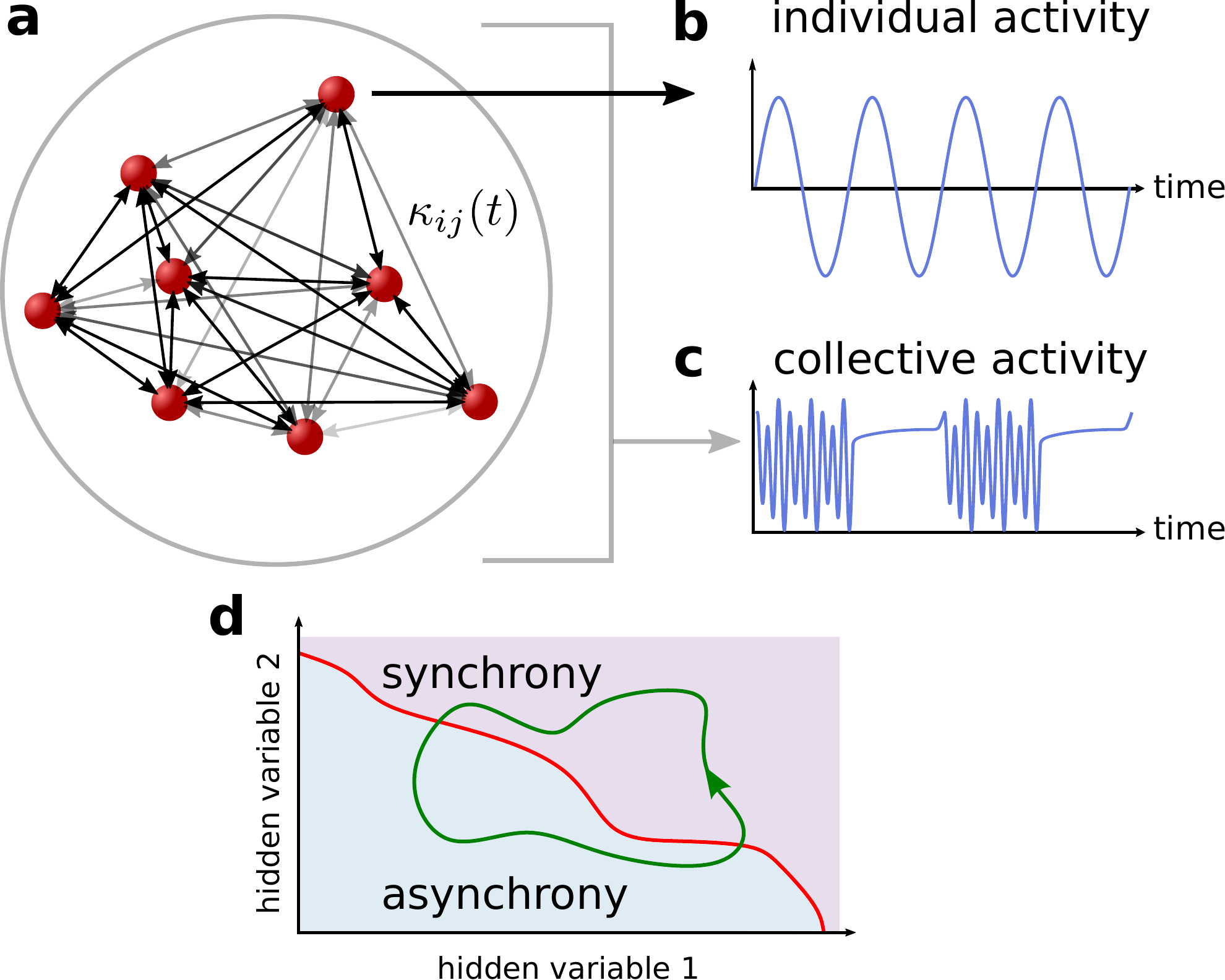}
    \caption{
    \textbf{Recurrent synchronization in a dynamical network.} 
    Panel \textbf{a} represents schematically a network of coupled nodes with time-dependent connections $\kappa_{ij}(t)$ (grayscale of the edges denotes the strength of the connection).
    Panel \textbf{b} displays a dynamical behavior of a single node which is periodic and does not exhibit bursting. 
    Panel \textbf{c} shows a bursting behavior of a macroscopic observable; it contains alternating episodes of steady and oscillatory dynamics. 
    Panel \textbf{d} shows how recurrent synchronization can be represented as a periodic orbit traversing through regions of synchronization and desynchronization with respect to hidden variables. 
    \label{fig:collective_bursting_schematic}
    }
\end{figure}

The reported phenomenon can be described in general terms, omitting nonessential details of the specific model implementation. The corresponding Fig.~\ref{fig:collective_bursting_schematic} summarizes the main phenomenology. We distinguish between the microscopic (individual) and the macroscopic (collective) scales. While the individual nodes show periodic oscillatory behavior (Fig.~\ref{fig:collective_bursting_schematic}b), the collective motion exhibits an alternation between time intervals of high activity, i.e., fast changes of the collective observable, and time intervals of low activity (Fig.~\ref{fig:collective_bursting_schematic}c), i.e., comparatively slow changes of the collective observable.

The mechanism for the recurrent intervals of synchronization and desynchronization can be explained by additional, hidden, slow variables (Fig.~\ref{fig:collective_bursting_schematic}d). 
Such hidden variables are not necessarily observable, but they play the key role as the internal control variables that determine the synchronization level.
As a result of a subtle interplay between the nodal dynamics 
and the hidden variables, alternating transitions between synchronized (phase-locked state) and desynchronized episodes (bursting state) can be induced (Fig.~\ref{fig:collective_bursting_schematic}d). As described below, the slow hidden variables naturally arise in networks with slow network adaptivity.

The recurrent synchronization phenomenon, which we report here, differs from those observed in \cite{BLA99b,STO02,GAS20,IZH00,MAR02a,ZEL18, PAN12, SCH08}, where single neurons possess alternating periods of spiking and rest. In our case, the nodes exhibit periodic behavior (tonic spiking) at all times, which makes the observed phenomenon even more surprizing.

\subsection{Recurrent synchronization in interacting populations of Hodgkin-Huxley neurons with spike-timing dependent plasticity}


Models of interacting populations are well-known paradigms for studying dynamics in many real-world systems on the mesoscopic scale \cite{KOM11,KOM13}. These models have particular importance in neuroscience for the modeling of interacting brain regions or other functional units~\cite{BAS18}; and are also used in social sciences with autonomous agents interacting with each other based on their population affiliation \cite{GON14,BEN16a}.

\begin{figure}
    \centering
    \includegraphics[width=0.95\textwidth]{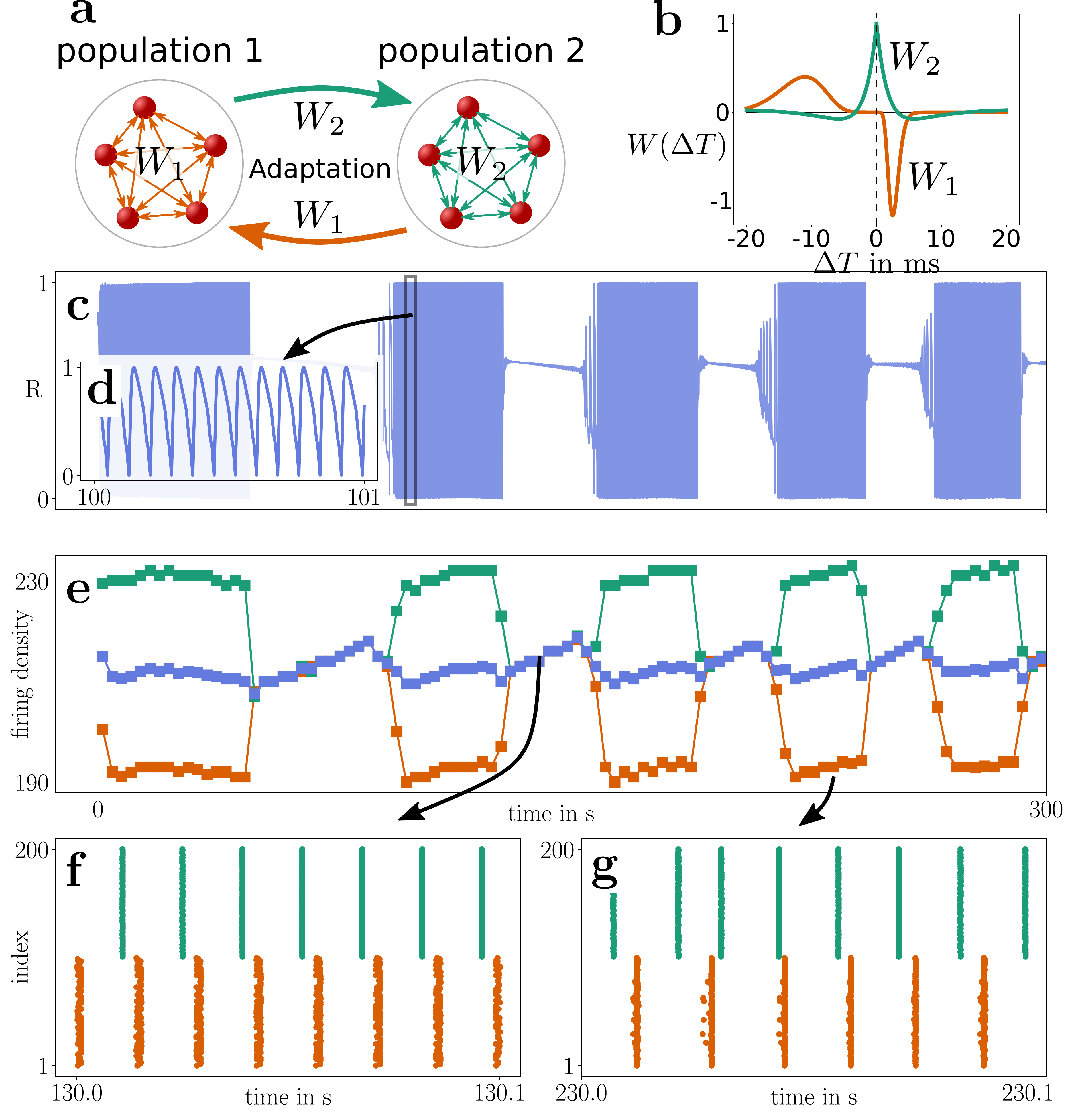}
    \caption{\textbf{Recurrent synchronization in a network of Hodgkin-Huxley neurons.} 
    Panel~\textbf{a} illustrates schematically the structure of the two-population network of Hodgkin-Huxley neurons with asymmetric spike-timing-dependent plasticity (shown in panel \textbf{b}). 
    The colors of the network links refer to the adaptation rule implemented within and between the populations. 
    Panel \textbf{c} shows the time series of the order parameter $R$ (blue, left scale) displaying multiple cycles of bursting and phase locked states. 
    Panel \textbf{d} shows a zoom into the bursting episode. 
    In panel \textbf{e}, the firing density is shown in blue for the whole network, in brown for population 1, and in green for population 2.
    Panels \textbf{f} and \textbf{g} display exemplary raster plots for the phase-locked and bursting episodes, respectively, with neurons of the first population (brown) and second population (green). 
    For simplicity, we consider an all-to-all coupling structure on which the coupling weights evolve. 
    The initial conditions for the coupling weights and the starting voltage were randomly chosen from the intervals $[0, 0.5]$ and $[-70\, \textnormal{mV}, 20\, \textnormal{mV}]$, respectively. 
    The constant input currents are randomly chosen from the intervals $[4.99\,\mu \mathrm{A},5.01\,\mu \mathrm{A}]$ and $[12.99\,\mu \mathrm{A},13.01\,\mu \mathrm{A}]$ for populations 1 and 2. All other parameters are provided in Section~\ref{sec:HHModel}.
    \label{fig:HHN_network}
    }
\end{figure}

To show recurrent synchronization in a complex dynamical network setup, we implement two populations of Hodgkin-Huxley neurons with spike-timing-dependent plasticity (STDP) (Fig. \ref{fig:HHN_network}), where the two populations differ in the adaptation functions of the coupling weights and in the input current (Fig.~\ref{fig:HHN_network}a). 
A detailed description of the plasticity rules, shown in Fig.~\ref{fig:HHN_network}b, and the Hodgkin-Huxley model is given in Section~\ref{sec:HHModel}.

Figure~\ref{fig:HHN_network}c and e show the temporal behavior of two macroscopic variables: the order parameter $R(t)$ which measures the level of synchronization, and the firing density measuring the mean level of activity of the whole system and the individual populations respectively.
The occurrence of several transitions between a nearly constant (phase-locking episode) to a strongly oscillating order parameter (bursting episode) is clearly visible. 
Figure~\ref{fig:HHN_network}d presents a zoom into a bursting episode, showing many oscillations in a small time interval of $1\,\mathrm{s}$. 
Figure~\ref{fig:HHN_network}e also displays the firing densities for the two different populations showing stark differences for the episodes of a fast oscillating order parameter. 
Figures~\ref{fig:HHN_network}f and~\ref{fig:HHN_network}g depict raster plots for the two dynamical episodes. 
It is clear that the collective bursting phenomenon in the order parameter is not due to the bursting behavior of individual neurons which show consistent tonic spiking at any time. The phenomenon is caused by frequency desynchronization between the two populations. 
The latter can be seen in the different inter-spike intervals of the populations. 
These results are also robust to noisy inputs, which we model with random $\alpha$-spikes, and higher heterogeneity in the input currents as well as heterogeneity in the update functions of the coupling weights, see Supplemental Note 7 for details.

Interestingly, a sharp drop in spiking activity coincides with both the onset and the termination of the bursting phase.


\subsection{Phenomenological mean field reduction}\label{sec:meanField}

The simulation in Fig.~\ref{fig:HHN_network} indicates that the dynamics within the populations is rather coherent while it can exhibit incoherence between the populations. 
In order to deepen the understanding of the emergence of recurrent synchronization, we propose the reduced model of the following general form:
\begin{eqnarray}
    \Dot{x}_1&=&F_1(x_1,x_2,\kappa_1),\label{eq:two_nodes-1}\\
    \Dot{x}_2&=&F_2(x_2,x_1,\kappa_2), \label{eq:two_nodes-2}\\
    \Dot{\kappa}_{1}&=&\epsilon W_1(x_1,x_2,\kappa_1), \label{eq:two_nodes-3}\\
    \Dot{\kappa}_{2}&=&\epsilon W_2(x_2,x_1,\kappa_2), \label{eq:two_nodes-4}
\end{eqnarray}
where the mean field variables $x_1$ and $x_2$ describe the coherent dynamics within the populations $1$ and $2$, respectively. 
The variable coupling weights $\kappa_1$ and $\kappa_2$ are the coupling weights representing the mean inter-population connections. In accordance with experimental findings~\cite{GER96}, we consider a small parameter $0<\epsilon\ll 1$, which enables the separation of time scales between the fast dynamics of the nodes and the slow dynamics of the coupling weights.
The adaptation dynamics is assumed to be on the same slow timescale $1/\epsilon$.
The functions $F_{1,2}$ govern the dynamical behavior of the corresponding populations and $W_{1,2}$ determine the adaptation rules of the coupling weights. 

Using the explicit splitting of the time scales (smallness of $\epsilon$), we develop a further reduction of the mean-field equations~\eqref{eq:two_nodes-1}-\eqref{eq:two_nodes-4}. In particular, as the coupling weights $(\kappa_1,\kappa_2)$ evolve slowly, we consider them as parameters for the dynamics of the oscillators. Depending on the coupling weights, the coupled oscillator system 
$\Dot{x}_1=F_1(x_1,x_2,\kappa_1)$, $\Dot{x}_2=F_2(x_2,x_1,\kappa_2)$ shows either synchronized or desynchronized motion. 

Denoting by $\langle \cdot \rangle$ the temporal averaging on the time scale which is much longer compared to the fast changes of the variables $x_{1,2}(t)$, but shorter than the timescale $1/\epsilon$ of the adaptation, we obtain $\langle W_{1}(x_1,x_2,\kappa_1)\rangle = \overline W_1(\kappa_1,\kappa_2)$, $\langle W_{2}(x_2,x_1,\kappa_2)\rangle = \overline W_2(\kappa_1,\kappa_2)$, and $\langle \dot \kappa_{1,2}\rangle \approx  \dot \kappa_{1,2}$. This leads to the following two-dimensional reduced model
\begin{equation}
\begin{aligned}
    {\kappa}'_{1}&= \overline W_1(\kappa_1,\kappa_2),\\
    {\kappa}'_{2}&= \overline W_2(\kappa_1,\kappa_2),
\end{aligned}
\label{eq:reduced}
\end{equation}
describing the dynamics of the coupling weights $\kappa_1$ and $\kappa_2$ on the slow time scale. 
In system \eqref{eq:reduced}, we have additionally rescaled time $t_s = \epsilon \cdot t$ and, thus, removed the factors $\epsilon$. The prime in $\kappa_1'$ and $ \kappa_2'$  denotes the derivative with respect to the new slow time $t_s$. The coupling weights $\kappa_1$ and $\kappa_2$ play the role of hidden variables governing the transitions between the synchronization and desynchronization as shown schematically in Fig.~\ref{fig:collective_bursting_schematic}d. System \eqref{eq:reduced} allows for a detailed bifurcation analysis of the mechanism behind the emergence of recurrent synchronization. Clearly, for recurrent synchronization, the dynamics of the hidden variables must exhibit recurrent motions. 
As shown in the following, such a recurrent motion can be observed when the adaptation functions between the two dynamical populations are non-symmetric.

As the system \eqref{eq:reduced} describes the (hidden) dynamics with respect to the slow time, it does not involve the small parameter $\epsilon$ and thus does not involve this time scale separation, and can be efficiently computed.

In Section~\ref{sec:methods}, we derive the models for slowly evolving coupling weights for two specific examples: two adaptively coupled phase oscillators and the more complex case of two coupled Hodgkin-Huxley neurons equipped with spike timing-dependent plasticity. The latter application shows, moreover, the generality of our approach with regards to complex local dynamics and adaptation rules.

From the mathematical point of view, the reduction is based on geometric singular perturbation theory~\cite{KUE15,DES12} and averaging~\cite{SAN07c} for the synchronized and desynchronized regimes, respectively. Both theoretical approaches are implemented analytically as well as semi-analytically. 


\subsection{Reduced Model: Two Hodgkin-Huxley neurons with spike-timing dependent plasticity}\label{sec:twoHHNeurons}

When the dynamical populations in Fig.~\ref{fig:HHN_network} are synchronized, the collective dynamics of each population can be approximated by a single Hodgkin-Huxley neuron. 
As a result, the dynamical network can be reduced to two coupled Hodgkin-Huxley neurons with asymmetric synaptic plasticity. 
The synaptic weight defined as $\kappa_1 = \kappa_{12}$ is updated accordingly to the STDP rule with the update function $W_1$, and $\kappa_2 = \kappa_{21}$ with $W_2$ (see Section \ref{sec:HHModel}), as in the case of two populations in Fig.~\ref{fig:HHN_network}.  

For such a system of two coupled Hodgkin-Huxley neurons with STDP, the reduction procedure from the previous Section \ref{sec:meanField} can be applied numerically. 
While the technical details are given in Sections~\ref{sec:adiabatic} and~\ref{sec:averaging}, here we present the main results.

\begin{figure}
    \centering
    \includegraphics[width=0.95\textwidth]{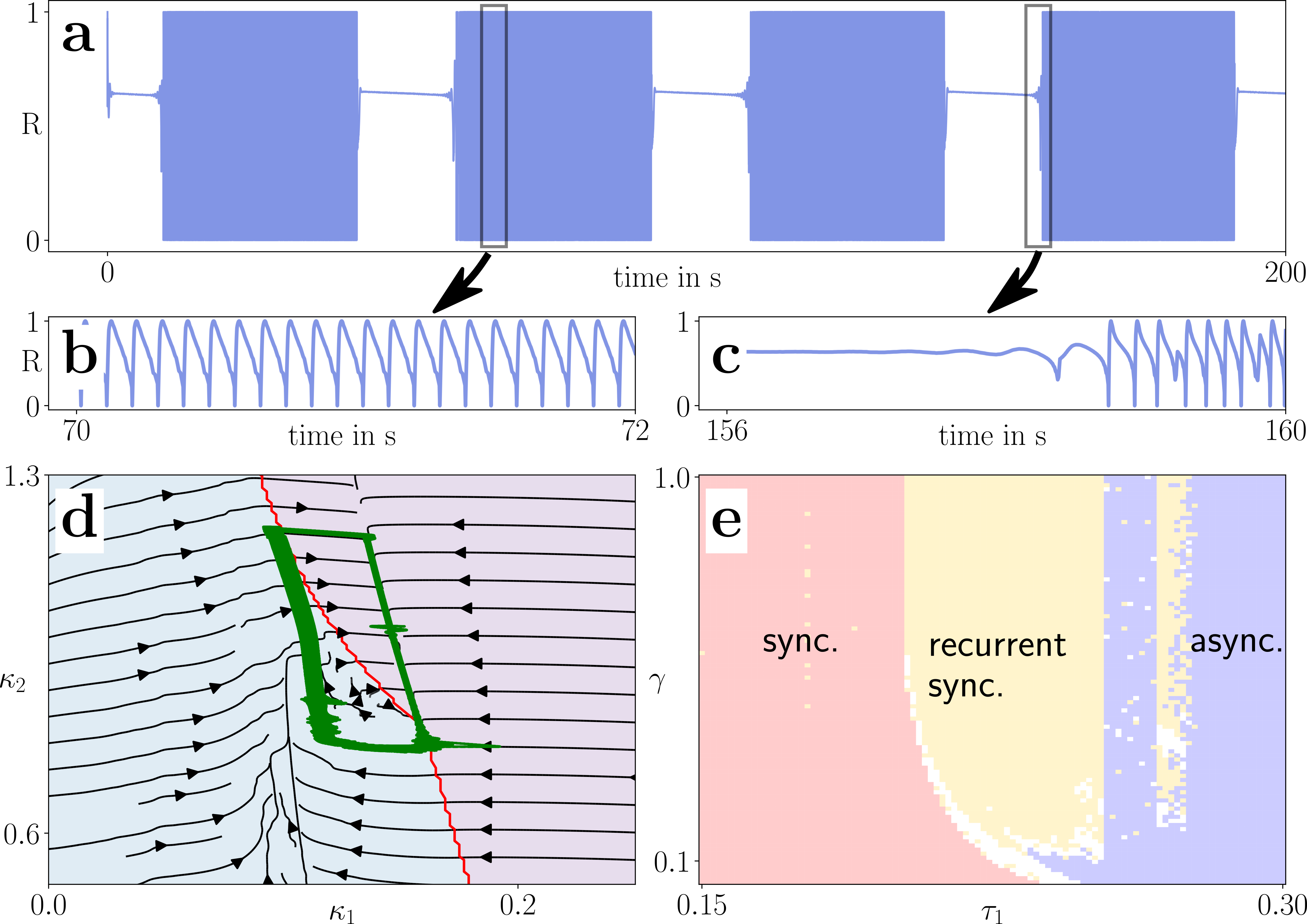}
    \caption{\textbf{Recurrent synchronization in a system of two Hodgkin-Huxley neurons with asymmetric synaptic plasticity.} 
    Panel~\textbf{a} shows a time series of the order parameter $R$, which
    exhibits recurrent synchronization.
    Panel~\textbf{b} and~\textbf{c} provide zooms into the time series of the order parameter for asynchronous spiking and at the transition from synchronous to asynchronous spiking, respectively.
    In panel~\textbf{d}, the phase portrait of the reduced system in coupling variables $(\kappa_1,\kappa_2)$ is shown. 
    Black lines show sample trajectories (flow lines) for the reduced system.
    The green curve is the projection of the whole system's trajectory onto the $(\kappa_1,\kappa_2)$-plane. 
    The red line marks the boundary between the synchronous (light pink) and asynchronous (light blue) regimes. 
    Panel~\textbf{e} shows three different asymptotic states identified numerically in the parameter plane $(\tau_1,\gamma)$ with trajectories starting in $(\kappa_1,\kappa_2)=(0.15,1)$. 
    Asynchronous spiking is marked in blue, recurrent synchronization is shown in yellow and phase locked spiking is depicted in red. The white areas indicate states which could not be categorized numerically.
    All parameters used for the simulation are given in Section~\ref{sec:HHModel}.
    \label{fig:2_HHN}
    }
\end{figure}

Figure~\ref{fig:2_HHN}a displays recurrent synchronization with a zoom into an asynchronous episode shown in Fig.~\ref{fig:2_HHN}b.
The reduced flow for the coupling weights \eqref{eq:reduced} is presented in Fig.~\ref{fig:2_HHN}d along with the projection of the simulated trajectory onto the $(\kappa_1,\kappa_2)$-plane. 
The recurrent synchronization, found in the simulations, corresponds to a periodic trajectory in the reduced model (green line). 
The trajectory in Fig.~\ref{fig:2_HHN}d enters the synchronous and asynchronous regimes recurrently. 
Moreover, the averaged flow (black lines) and the projected flow based on the simulation of the whole system are in very good agreement, and only small deviations near the boundary can be observed. 
Therefore, our proposed reduction method which eliminates the fast dynamics of the oscillators while keeping the slow dynamics of the coupling weights, can also be applied to discontinuous adaptation functions. 

A closer view into the transition from synchronization to desynchronization can be seen in Figure \ref{fig:2_HHN}c. At the end of a synchronized episode, the oscillations of the order parameter are gradually building up from small to large variations. This observation could be a useful indicator for the onset of bursting events where small variations of the order parameter may be interpreted as precursors.

A numerical bifurcation diagram for parameters $\tau_1$ (parameter of $W_1$) and $\gamma$ (parameter of $W_2$) is presented in Fig.~\ref{fig:2_HHN}e. The parameter regions for which recurrent synchronization is observed are shown in yellow. The emergence of recurrent synchronization for different parameter regimes is clearly visible, with even two unconnected parametric regions. Therefore, the phenomenon of recurrent synchronization is observed for a wide range of parameters. The periodic motion in the hidden variables shown in Fig.~\ref{fig:2_HHN}d is induced by asymmetric adaptation (see Fig.~\ref{fig:HHN_network}b) and is not be found for symmetric adaptation rules. For comparison, in Supplemental Note 6 we show the dynamics in case of symmetric adaptation.

\subsection{Reduced Model: Two Phase Oscillators}\label{sec:twoPhaseOsc}

In order to obtain deeper insight into recurrent synchronization, we consider an even simpler paradigmatic model for the dynamics of the synchronous population: the phase oscillator \cite{PIK01,KUR88}. 
Taking into account the adaptation, the model of two adaptively coupled phase oscillators  reads~\cite{KAS17,BER19} 
\begin{eqnarray}
    \dot{\phi}_1&=&\omega_1-\kappa_1\sin(\phi_1-\phi_2+\alpha), \label{eq:2_PO-1}\\
    \Dot{\phi}_2&=&\omega_2-\kappa_2\sin(\phi_2-\phi_1+\alpha), \label{eq:2_PO-2}\\
    \Dot{\kappa}_{1}&=&-\epsilon(\kappa_{1} - a\sin(\phi_1-\phi_2)), \label{eq:2_PO-3}\\
    \Dot{\kappa}_{2}&=&-\epsilon(\kappa_{2} - b\sin(\phi_2-\phi_1+\beta)), \label{eq:2_PO-4}
\end{eqnarray}
where $\phi_i \in [0,2\pi)$ represents the phases of the oscillatory dynamics for each population. When the phase increases from $0$ to $2\pi$, this corresponds to one oscillation period. The natural frequencies of the $i$th oscillator are denoted by $\omega_i$. The adaptive coupling weights are given by $\kappa_{i}$. 
The phase lag parameter $\alpha$ is included that may account for a small information transmission delay~\cite{ASL18a}. 
The parameters $a$ and $b$ scale the influence of the phase dynamics on the coupling weights. 
An additional asymmetry in the adaption rules for $\kappa_1$ and $\kappa_2$ is introduced by the parameter $\beta$. 

Applying the averaging procedure introduced in Sec.~\ref{sec:meanField},    system~\eqref{eq:2_PO-1}--\eqref{eq:2_PO-4} is reduced to the following two-dimensional model
\begin{equation}\label{eq:twoPhaseOsc_red}
\begin{aligned}
    {\kappa}'_{1}&=- \kappa_{1} + a\widehat{W}_1(\kappa_{1},\kappa_{2}), \\
    {\kappa}'_{2}&=- \kappa_{2} + b\widehat{W}_2(\kappa_{1},\kappa_{2}),
\end{aligned}
\end{equation}
where $\widehat{W}_{1}( \kappa_1,\kappa_2)=\langle \sin (\phi_1 (t)-\phi_2 (t)) \rangle$ and $\widehat{W}_{2}( \kappa_1,\kappa_2)=\langle \sin (\phi_2 (t)-\phi_1 (t) + \beta) \rangle$ are the corresponding temporal averages of the adaptation functions.
We emphasize that the explicit dependence of the functions $\widehat{W}_{1,2}$ on the coupling weights stem from the dependence of the fast dynamics $\phi_1(t)$ and $\phi_2(t)$ on these weights. 



The functions $\widehat{W}_{1,2}( \kappa_1,\kappa_2)$ are computed explicitly in Sections~\ref{sec:adiabatic} and~\ref{sec:averaging}, which allows the properties of the reduced system \eqref{eq:twoPhaseOsc_red} to be studied in detail. 
While the technical details of the reduction are described in Sections~\ref{sec:adiabatic} and~\ref{sec:averaging}, we note here that this procedure is performed by two different methods: adiabatic elimination for such parameter values $\kappa_1$, $\kappa_2$ that the subsystem \eqref{eq:2_PO-1}--\eqref{eq:2_PO-2} is synchronized (phase-locked) and the averaging along a periodic orbit when \eqref{eq:2_PO-1}--\eqref{eq:2_PO-2} is not synchronized.

\begin{figure}
    \includegraphics[width=\textwidth]{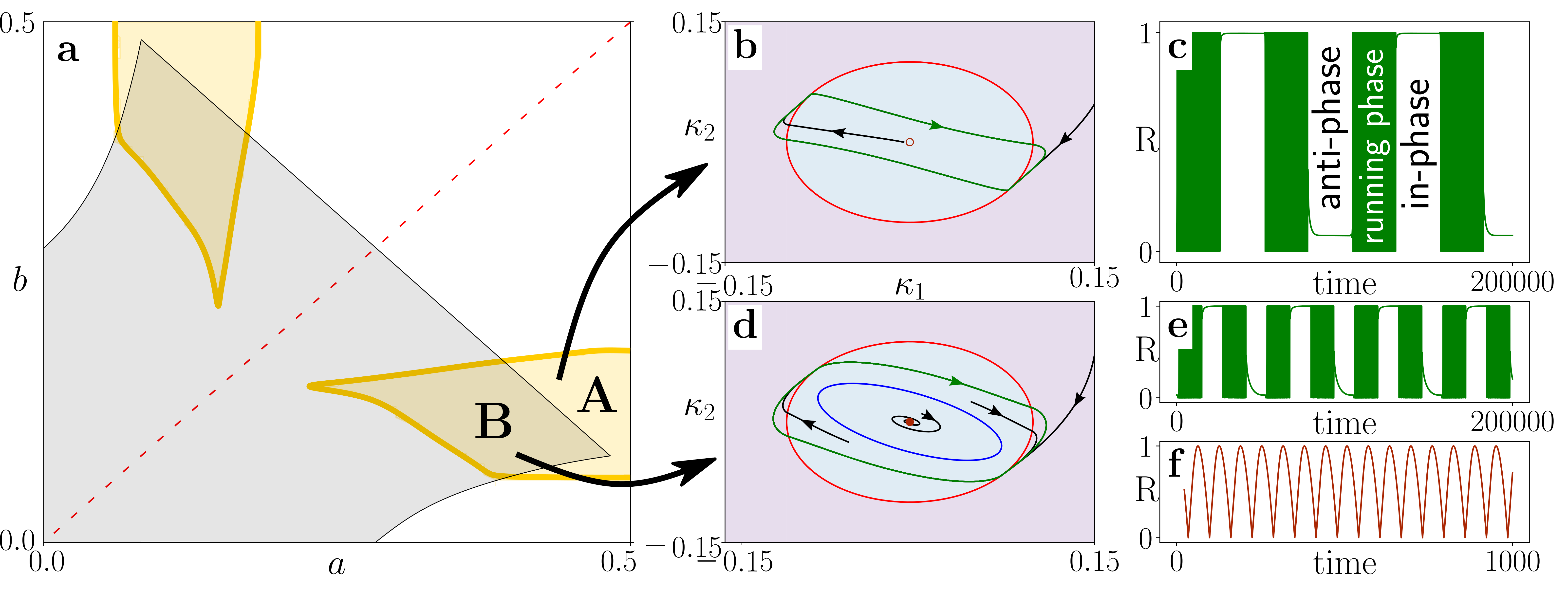}
    \caption{
    \textbf{Recurrent synchronization in the phase oscillator model~\eqref{eq:2_PO-1}--\eqref{eq:2_PO-4}. }
    In \textbf{a}, the bifurcation diagram in the $(a,b)$-parameter plane is shown. 
    A and B denote the regions with stable recurrent synchronization, i.e., the stable limit cycle traversing the synchronous and asychronous domain in $(\kappa_1,\kappa_2)$ plane. 
    Region B corresponds to the coexistence of the bursting cycle and the stable equilibrium at $\kappa_1=\kappa_2=0$. 
    Panels \textbf{b} and \textbf{d} show the trajectories of the reduced system \eqref{eq:twoPhaseOsc_red} in the $(\kappa_1,\kappa_2)$-plane that correspond to the time series in \textbf{c} and \textbf{e,f}, respectively.
    The phase portrait \textbf{d} is complemented by an additional trajectory (blue) separating the basins of attraction of the recurrent synchronization limit cycle and the stable equilibrium at $(\kappa_1,\kappa_2)=(0,0)$. 
    The red line denotes the boundary between the asynchronous (gray blue) and the phase locked (light pink) regions.
    Panels \textbf{c} ($a=0.5$, $b=0.07$) and \textbf{e, f} ($a=0.385$, $b=0.125$) show the time evolution of the order parameter $R$ for the recurrent synchronization (green) and the trajectory converging to the equilibrium at $(0,0)$ (dark red). 
    Parameters: $\epsilon=0.0001$, $\omega=0.1$, $\alpha=0.25\pi$, $\beta=-\pi/2$.}
    \label{fig:2_phase_osc}
\end{figure}

The bifurcation diagram for system~\eqref{eq:twoPhaseOsc_red} in Fig.~\ref{fig:2_phase_osc} shows the regions for which recurrent synchronization is observed (yellow) in the $(a,b)$-parameter plane for fixed parameters $\omega=\omega_1-\omega_2$ and $\alpha$.  
In the parameter region A, the recurrent synchronization is the only stable regime (Fig.~\ref{fig:2_phase_osc}b), while in region B recurrent synchronization coexists with an effectively uncoupled state, i.e., equilibrium at $(\kappa_{1},\kappa_{2})=(0,0)$ of the reduced system (Fig.~\ref{fig:2_phase_osc}d). 

Figures~\ref{fig:2_phase_osc}c and \ref{fig:2_phase_osc}e show the time evolution of the order parameter~\eqref{eq:order_par} in the regime of recurrent synchronization. Interestingly, we observe the repeated occurrence of two synchronous regimes: an in-phase and an anti-phase locking interrupted with the running phase regime. 
In Figure \ref{fig:2_phase_osc}f, yet another stable state (red line) is shown corresponding to desynchronization of the two oscillators leading to an effective decoupling. 
The corresponding states in the reduced model are displayed in the phase portraits in Figs. \ref{fig:2_phase_osc}b and \ref{fig:2_phase_osc}d. 
These phase portraits of~\eqref{eq:twoPhaseOsc_red} show limit cycles clearly passing through regions of synchronous and asynchronous relative phase dynamics, which explains the dynamics of the order parameter in panels Fig.~\ref{fig:2_phase_osc}c and e.

The bifurcation analysis reveals that the emergence of recurrent synchronization is the result of a fold bifurcation of the limit cycle of the reduced system. For more details, see Supplemental Note~3.
Moreover, the bifurcation diagram shows that recurrent synchronization can be observed for a wide range with respect to the adaptation parameters $a$ and $b$. 
Beyond this, recurrent synchronization can also be found for other degrees of asymmetry, i.e., other values of the asymmetry parameter $\beta$, see Supplemental Note~6 and multiple orders of time scale separation, see Supplemental Note~5. 
In particular, the bifurcation diagram for the symmetric case $\beta=0$ reveals no recurrent synchronization. In addition to the need for asymmetry in the shape of the plasticity rules, non-identical scaling of the plasticity function is also crucial for the occurrence of recurrent synchronization, i.e., $|a|\ne|b|$. Therefore, the regions for recurrent synchronization lie off the lines $|a|=|b|$ in Fig.~\ref{fig:2_phase_osc}a. 

\section{Discussion}\label{sec:discussion}

In this work, we have described recurrent synchronization for two populations of adaptively coupled oscillators. 
The phenomenon is shown to be robust to noise, variation of parameters, initial states, and heterogeneity of the oscillators.

To shed light on the emergence of recurrent synchronization, we have introduced a mean-field model approach. 
A detailed bifurcation analysis of a simplified phenomenological phase oscillator model provides remarkable insights into the mechanisms leading to recurrent synchronization.
The separation between the time scales of the adaptation and the fast individual dynamics allows for the model reduction in the case of two phase oscillators and two Hodgkin-Huxley neurons. 
The proposed reduction approach is applicable to systems beyond phase oscillators and, in this regard, it methodologically generalizes recently used methods from Ratas et. al.~\cite{RAT21} and Franovi\'{c} et. al.~\cite{FRA20}. 
Our study is further complemented by an analysis of a mean-field model consisting of coupled Hodgkin-Huxley neurons equipped with spike-timing-dependent plasticity. For this more complicated system, our numerical averaging approach makes the study of the effects of plasticity on neuronal dynamics much more feasible.

Heterogeneity in the adaptation rule has been known to exist e.g. in neural systems for a long time~\cite{LET06} but their dynamical effects are only rarely studied and analytical methods are widely unexplored~\cite{KAS21a,BER21f}. 
Our findings contribute to the important question how heterogeneity may change the dynamics of complex networks. We have found that such heterogeneity in the adaptivity is able to significantly change the macroscopic dynamics of a dynamical network. Moreover, our analysis contributes to the identification of the onset of critical phenomena and extreme events. In the transition from synchrony to asynchrony, we have observed gradually growing oscillations of the order parameter. This observation may be used as characterizing feature for the onset of the bursting period and thus may indicate substantial changes in the properties of the network. 

Our results may contribute to the understanding of pattern generators, their disease-related impairment and, specifically, the origin of Parkinsonian resting tremor. So far, the mechanism underlying the generation of Parkinsonian resting tremor remains an open question~\cite{RAE00,BEN01,HEL17}. Neuronal discharges in the tremor frequency range in different parts of the basal ganglia as well as the thalamus were found to be coherent and transiently locked to the peripheral tremor~\cite{BRO03a}. Furthermore, causal data analysis suggests that this activity drives the muscular tremor, as opposed to being just driven by the sensory feedback from the periphery~\cite{TAS10}. In functional magnetic resonance imaging (fMRI) studies it was shown that basal ganglia get active transiently when tremor episodes emerge, whereas activity in the cerebello-thalamo-cortical circuit is tightly related to tremor amplitude~\cite{HEL17}. According to the ”dimmer-switch” hypothesis, the basal ganglia activate the tremor (like a light switch) and the cerebello-thalamo-cortical circuit modulates tremor amplitude (like a light dimmer)~\cite{HEL12,HEL17}. However, it remains elusive what causes spontaneous switching between epochs with alternating (anti-phase) tremor and synchronous (in-phase) tremor, connected by epochs without stable phase relationship~\cite{BIS48,RAE00}. Based on our results, we suggest that asymmetric adaptivity involving two populations (Fig.~\ref{fig:HHN_network}) responsible for two antagonistic muscles might cause the clinically observed recurrent epochs of synchrony with different phase differences, e.g., in-phase and anti-phase tremor epochs. In that sense, asymmetric adaptivity might serve as a complex dynamical switching mechanism. In contrast, two anatomically intermingled populations (Fig.~\ref{fig:collective_bursting_schematic}) activating the same muscle would display recurrent epochs of synchrony, interconnected by bursts of mass synchrony, potentially translating to pronounced tremor epochs, intersected by epochs of, e.g., less coherent twitches, lacking distinct tremor bursts.

Bursting, as it has been discussed in the literature e.g. for individual neurons, is a phenomenon consisting of episodes of activity and inactivity~\cite{IZH00}. 
On the other hand, recurrent synchronization, as it has been introduced in this work, is a phenomenon arising from the interaction of oscillators in a complex dynamical network. 
It is characterized by bursting on the macroscopic level rather than on the level of individual microscopic dynamical units. 
  
While this work is exemplified by using the Hodgkin-Huxley model to show the generality of our findings, the theory developed is not restricted to the field neuroscience. Moreover, we note that the model used in this work represents a phenomenological model for a complex dynamical system inspired by neuroscience rather than a realistic network model.
As we show, recurrent synchronization is a result of the interplay of a coupling structure with the adaptation on a slower time scale.
The "hidden" adaptation variables play the role of the slowly changing control parameters triggering the recurrent synchronization events. 
Recurrent synchronization induced by the interplay of multiple timescales and complex dynamical networks is inherent in many dynamical networks modeling, e.g., social interactions or climate tipping dynamics \cite{ASH17,HEY21}. 
Bursting phenomena are also known dynamical states found outside the field of neuroscience~\cite{TAN20a}. 


\section{Methods}\label{sec:methods}

\subsection{Network of Hodgkin-Huxley neurons}\label{sec:HHModel}
The dynamics for the network of synaptically coupled Hodgkin-Huxley neurons are given by the following equations \cite{HOD52,HAN93b}:
\begin{equation}
\begin{aligned}
    C \dot{V}_{i} &=I_{i}-g_{N a} m_{i}^{3} h_{i}\left(V_{i}-E_{N a}\right)-g_{K} n_{i}^{4}\left(V_{i}-E_{K}\right)-g_{L}\left(V_{i}-E_{L}\right)-\frac{\left(V_{i}-E_{r}\right)}{N} \sum_{j=1}^{N} \kappa_{i j} s_{j}, \\
    \dot{m}_{i} &=\alpha_{m}\left(V_{i}\right)\left(1-m_{i}\right)-\beta_{m}\left(V_{i}\right) m_{i}, \\
    \dot{h}_{i} &=\alpha_{h}\left(V_{i}\right)\left(1-h_{i}\right)-\beta_{h}\left(V_{i}\right) h_{i}, \\
    \dot{n}_{i} &=\alpha_{n}\left(V_{i}\right)\left(1-n_{i}\right)-\beta_{n}\left(V_{i}\right) n_{i}, \\
    \dot{s}_{i} &=\frac{5\left(1-s_{i}\right)}{1+e^{\left(\frac{-V_{i}+3}{8}\right)}}-s_{i}.
\end{aligned}
\end{equation}
Here $V_{i}$ is the membrane potential of the $i$th neuron, $E_{N a}=50\,\mathrm{mV}$, $E_{K}=-77\,\mathrm{mV}$, and $E_{L}=-54.4\,\mathrm{mV}$ are reverse potentials of the different ion channels. 
The dynamic variables $m_i$, $h_i$ and $n_i$ denote gating variables describing the probability of open ion-channels for sodium activation, sodium deactivation and potassium, respectively.
The membrane capacity is given by $C=1\,\mu\mathrm{F} / \mathrm{cm}^{2}$. We set the reverse potential of the coupling $E_{r}=20\,\mathrm{mV}$ which corresponds to excitatory neurons.
The functions $\alpha_i$ and $\beta_i$ are given as
\begin{equation*}
\begin{aligned}
    &\alpha_{m}(V)=\frac{0.1 V+4}{1-e^{(-0.1 V-4)}}, \\
    &\beta_{m}(V)=4 e^{\left(\frac{-V-65}{18}\right)}, \\
    &\alpha_{h}(V)=0.07 e^{\left(\frac{-V-65}{20}\right)}, \\
    &\beta_{h}(V)=\frac{1}{1+e^{(-0.2 V-3.5)}}, \\
    &\alpha_{n}(V)=\frac{0.01 V+0.55}{1-e^{(-0.1 V-5.5)}}, \\
    &\beta_{n}(V)=0.125 e^{\left(\frac{-V-65}{80}\right)} .
\end{aligned}
\end{equation*}
The corresponding conductance parameters are set to  $g_{Na}=120\,\mathrm{mS} / \mathrm{cm}^{2}$, $g_{K}=36\,\mathrm{mS} / \mathrm{cm}^{2}$, and $g_{L}=0.3\,\mathrm{mS} / \mathrm{cm}^{2}$. 
The constant input currents are randomly chosen from the intervals $[4.99\,\mu \mathrm{A},5.01\,\mu \mathrm{A}]$ and $[12.99\,\mu \mathrm{A},13.01\,\mu \mathrm{A}]$ for populations 1 and 2, respectively.
Different input currents $I_i$ account for different spiking rates of the two populations and different neurons.

The coupling terms include the synaptic activity $s_j$ of the presynaptic neuron, multiplied by the coupling weights $\kappa_{ij}$, which changes discontinuously due to synaptic plasticity every time the $i$th or $j$th neuron exhibits a spike. 
Whenever a neuron spikes, the coupling weights between the neurons receive an update according to
$\kappa_{ij} \mapsto \kappa_{ij}+\delta\cdot  W\left(\Delta T\right)$ ($\delta=0.005$) which depends on the timing difference $\Delta T_{ij}=t_i-t_j$ between the last spiking events $t_i$ and $t_j$ of the $i$th and $j$th neuron, respectively. We identify a spiking event by checking whether the voltage of the neuron in question exceeds 0 in the given time step.
For the two populations, we use different update functions $W$. 
For population 1 (coupling weights of all connections with postsynaptic neurons in population 1),  the update function is anti-Hebbian~\cite{KIM20} and of the following form:
\begin{equation}
W_1^\textnormal{STDP} =
\begin{cases}
-A_1\,\mathrm{exp}\left (-\frac{\Delta T_{ij} }{\tau_1} \right)\left(\frac{\Delta T_{ij}e}{10\tau_1}\right)^{10}, & \text { if } \Delta T_{ij}>0 \\
A_2\,\mathrm{exp}\left (\frac{\Delta T_{ij} }{\tau_2} \right)\left(\frac{\Delta T_{ij}e}{10\tau_2}\right)^{10}, & \text { if } \Delta T_{ij}<0 \\
0, & \text { if } \Delta T_{ij}=0.
\end{cases}
\end{equation}
The parameters used for Fig. \ref{fig:HHN_network} and Fig. \ref{fig:2_HHN} are $A_1=1.17$, $A_2=0.4$, $\tau_1=0.25$, $\tau_2=1.1$ and $e=\exp(1)$.
For population 2, the update function is symmetric \cite{ROE19a} and of the following form:
\begin{equation}
    W_2^\textnormal{STDP}\left(\Delta T_{ij}\right)=\gamma\left(c_{p} e^{-\frac{\left|\Delta T_{ij}\right|}{\tau_{p}}}-c_{d} e^{-\frac{\left| \Delta T_{ij} \right|}{\tau_{d}}}+\frac{1}{30}\right)
\end{equation}
with $c_p=1.5$, $c_d=0.53$, $\tau_p=1.8$, and $\tau_d=5$. 
To ensure bounded coupling weights, we restrict them to the interval $[0,1.5]$. The two plasticity rules exemplify the diversity of adaptation rules known for nervous systems. We note, however, that the dynamical system introduced in this work can be conceived as a toy model possessing complex dynamical features rather than a model of a realistic system.

\subsection{Macroscopic Observables}\label{sec:macrObs}

To measure the collective motion of the whole network, we introduce an observable that provides an average of the individual node dynamics. 
The behavior of each dynamical node $x_i(t)$ can be mapped to a motion on the unit circle represented by a phase variable $\phi_i(t)$ \cite{PIK01}.
Particularly, for Hodgkin-Huxley neurons considered in Section~\ref{sec:results} each node possesses a sequence of spiking events $\{t_{i,1}, t{_{i,2}},\dots\}$. 
The mapping used in this study is given by
\begin{align*}
    \phi_i(t) = 2\pi\frac{t-t_{i,k}}{t_{i,k+1}-t_{i,k}}
\end{align*}
for $t_{i,k}\le t<t_{i,k+1}$. The collective observable is then defined by the Kuramoto order parameter
\begin{align}\label{eq:order_par}
    R(t) = \frac{1}{N}\left|\sum_{j=1}^N \exp(\mathrm{i}\phi_j(t))\right|.
\end{align}
If, for a certain time $t$, the phases are incoherently spread over the interval $[0,2\pi)$, the order parameter is zero while it is unity if all phases are the same. 
Further, by looking at the temporal behavior of $R(t)$, we are able to distinguish between phase-locking, which corresponds to an order parameter approximately constant in time, i.e., $R(t)\approx \mathrm{const}$, and the loss of phase relation which corresponds to an order parameter oscillating between incoherence (small values of $R$) and coherence $R\approx 1$.

For the network of Hodgkin-Huxley neurons, we consider the firing density as a second macroscopic observable. The firing density is determined by dividing the time period of observation into time bins of width $=3000\,\mathrm{ms}$ and counting the spikes for each time bin. Afterwards, the count per time bin is divided by the number of neurons in the group.

\subsection{Adiabatic elimination}\label{sec:adiabatic}

To obtain the reduced two-dimensional flow \eqref{eq:reduced} for system \eqref{eq:two_nodes-1}-\eqref{eq:two_nodes-4}, we average it over a time interval, which is much larger than the fast timescale of the subsystem~\eqref{eq:two_nodes-1}-\eqref{eq:two_nodes-2}, but smaller than $1/\epsilon$. 
Such an average is easily calculated for the case when the dynamics of the subsystem \eqref{eq:two_nodes-1}-\eqref{eq:two_nodes-2}  (with fixed $\kappa_1$ and $\kappa_2$) converges to an equilibrium $(x_1^*(\kappa_1,\kappa_2),x_2^*(\kappa_1,\kappa_2))$. 
The procedure is known as "adiabatic elimination", and the subsystem \eqref{eq:two_nodes-1}-\eqref{eq:two_nodes-2} as the "layer equation" or "fast subsystem" \cite{KUE15}.
If the fast subsystem has one or more equilibria, then they satisfy the equations:
\begin{equation}
    \begin{aligned}
        0&=F_1\left(x_{1,i}^*(\kappa_1,\kappa_2),x_{2,i}^*(\kappa_1,\kappa_2),\kappa_1\right), \\
        0&=F_2\left(x_{2,i}^*(\kappa_1,\kappa_2),x_{1,i}^*(\kappa_1,\kappa_2),\kappa_2\right), 
    \end{aligned}
\label{eq:CM}
\end{equation}
where $x_{1,i}^*$ and $x_{2,i}^*$ denotes the \textit{i}-th solution and the union of all solutions ($x_1^*$, $x_2^*$) compose the so called critical manifold. 
If an equilibrium $(x_{1,i}^*(\kappa_1,\kappa_2),x_{2,i}^*(\kappa_1,\kappa_2)$ is stable with respect to the fast subsystem \eqref{eq:two_nodes-1}-\eqref{eq:two_nodes-2}, then the solutions will rapidly converge to this equilibrium, and the reduced system \eqref{eq:reduced} can be simply obtained by substituting this equilibrium point into \eqref{eq:two_nodes-3}-\eqref{eq:two_nodes-4}:
\begin{equation}
\label{eq:reduced-1}
    \begin{aligned}
        \kappa_1'&=  \overline W_1(\kappa_1,\kappa_2) =  W_1\left(x_{1,s}^*(\kappa_1,\kappa_2),x_{2,s}^*(\kappa_1,\kappa_2),\kappa_1\right) , \\
        \kappa_2'&=  \overline W_2(\kappa_1,\kappa_2) =  W_2\left(x_{1,s}^*(\kappa_1,\kappa_2),x_{2,s}^*(\kappa_1,\kappa_2),\kappa_2\right).
    \end{aligned}
\end{equation}
For the system of two phase oscillators \eqref{eq:2_PO-1}-\eqref{eq:2_PO-4}, the adaptation functions are $W_1=a\sin\theta-\kappa_1$, $W_2=-b\cos\theta-\kappa_2$, and this yields the following explicit form for the reduced system:
\begin{equation}
\label{eq:reduced-phase}
\begin{aligned}
    \kappa_1'&=- \kappa_1 + a\sin\left(\arcsin\frac{\omega}{\sqrt{c_1^2+c_2^2}}-\arg(c_1+\mathrm{i}c_2)\right),\\
    \kappa_2'&= - \kappa_2 - b\cos\left(\arcsin\frac{\omega}{\sqrt{c_1^2+c_2^2}}-\arg(c_1+\mathrm{i}c_2)\right),
\end{aligned}
\end{equation}
with 
\begin{equation}
\label{eq:c}
 c_1=\left(\kappa_1+\kappa_2\right)\cos\alpha,\quad
 c_2=\left(\kappa_1-\kappa_2\right)\sin\alpha.
\end{equation}
A more detailed derivation of this equation is given in the Supplemental Note 1. 

To obtain the reduced system for two Hodgkin-Huxley neurons from Section~\ref{sec:twoHHNeurons} in the synchronized regime, we use direct averaging (with fixed $\kappa_1$ and $\kappa_2$) as described in Section \ref{sec:averaging}, since the synchronized solution corresponds to an oscillatory regime rather than an equilibrium in that case.

\subsection{Averaging fast oscillations}\label{sec:averaging}

If the fast subsystem \eqref{eq:two_nodes-1}-\eqref{eq:two_nodes-2} does not converge to a stable equilibrium, the adiabatic elimination described in \ref{sec:adiabatic} cannot be applied. 
In such a case, direct averaging must be used. 
If, for fixed $(\kappa_1,\kappa_2)$, the solution of \eqref{eq:two_nodes-1}-\eqref{eq:two_nodes-2} converges to a periodic state $x_1(t;\kappa_1,\kappa_2)$, $x_2(t;\kappa_1,\kappa_2)$ with a period $T(\kappa_1,\kappa_2)$, then this averaging procedure leads to 
\begin{equation}
\label{eq:aver-periodic}
    \begin{aligned}
        {\kappa}'_1 & = \frac{1}{T}\int_0^T W_1\left(x_1(t),x_2(t),\kappa_1 \right) dt,\\
        {\kappa}'_2  &=\frac{1}{T}\int_0^T W_2\left(x_2(t),x_2(t),\kappa_2 \right) dt.
    \end{aligned}
\end{equation}
For the phase oscillators \eqref{eq:2_PO-1}-\eqref{eq:2_PO-2}, the asynchronous dynamics is always periodic, and the above integrals can be explicitly found, leading to the following reduced averaged system
\begin{equation}
    \begin{aligned}
            {\kappa}'_1= + \frac{c_1 a}{c_1^2+c_2^2}\left(\omega-\sqrt{\omega^{2}-c_1^2-c_2^2}\right)-\kappa_1, \\
            {\kappa}'_2 =-\frac{c_2 b}{c_1^2+c_2^2}\left(\omega-\sqrt{\omega^{2}-c_1^2-c_2^2}\right)-\kappa_2,
    \end{aligned}
\end{equation}
where $c_1$ and $c_2$ are given in Eq.~\eqref{eq:c}, see Supplemental Note 2 for details.
The averaging approximation \eqref{eq:aver-periodic} is also applicable to two Hodgkin-Huxley neurons when the dynamics is periodic. 
For more complicated dynamics, the averaging in \eqref{eq:aver-periodic} is taken for sufficiently large $T$.
Practically, the slow flow for the two Hodgkin-Huxley neurons in Fig.~\ref{fig:2_HHN}b is obtained using a grid in the plane of the coupling weights $(\kappa_1,\kappa_2)$ and estimating the averages in \eqref{eq:aver-periodic} for each grid point.  
For this, the distribution $\rho(\xi)$ of inter-spike intervals $\Delta T$ is first computed. 
Then the right-hand side of the reduced systems are estimated as 
\begin{equation}
\label{eq:stdp-averaging}
\overline W_i (\kappa_1,\kappa_2) = \int W_i^\textnormal{STDP}(\xi) \rho(\xi) d\xi. 
\end{equation}
Note that any scaling factors in \eqref{eq:stdp-averaging}, including $\delta$, can be ignored as they can be scaled out by the time rescaling, similarly to the rescaling of the small parameter $\epsilon$.

Importantly, the fast systems have to be computed once in order to find the distributions $\rho(\xi)$ for every grid point in $\kappa_1,\kappa_2$ plane. 
Once these data are created, the slow flow can be computed using \eqref{eq:stdp-averaging} for any update function without simulating the fast system. 
This makes the study of the effects of different plasticity rules much more feasible.

\section*{Acknowledgments}
This work was supported by the German Research Foundation DFG, Projects No. 411803875 and No. 440145547.

\section*{Author contributions}
R.B. and S.Y. designed and supervised the research. M.T. performed the numerical and theoretical analysis. M.T., R.B., P.A.T., E.S. and S.Y. conceptualized the manuscript. M.T., R.B. and S.Y. prepared the manuscript. M.T., R.B., P.A.T., E.S. and S.Y. interpreted the results, drew conclusions and edited the manuscript.

\section*{Code availability}
The source code to reproduce the results of this study is freely available on GitHub:\newline
\url{https://github.com/maxthiele/RecurrentSynchronization.git}.

\printbibliography

\title{Supplemental Material}

\renewcommand{\thefigure}{S.\arabic{figure}}
\renewcommand{\theequation}{S.\arabic{equation}}
\renewcommand{\thesection}{Supplemental note \arabic{section}:}
\setcounter{figure}{0} 
\setcounter{section}{0} 
\setcounter{equation}{0}

\maketitle

\section{Adiabatic elimination for two phase \\ oscillators}

In section 4.3 of the main text, we introduced the concept of adiabatic elimination and showed the resulting flow for system (6)--(9) of the main text. 
In the following, we provide a derivation of this flow for the following adaptation functions: $W_1=a\sin\theta-\kappa_1$ and $W_2=b\cos\theta-\kappa_2$. 
We start with reducing the model (6)--(9) of the main text by introducing the phase difference $\theta=\phi_1-\phi_2$ and the frequency difference $\omega=\omega_1-\omega_2$:
\begin{subequations}
\begin{alignat}{3}
    \Dot{\theta}&=\omega-\kappa_1\sin(\theta+\alpha)-\kappa_2\sin(\theta-\alpha),\label{eq:2_PO_theta_1} \\
    \Dot{\kappa}_{1}&=-\epsilon(-a\sin\theta+\kappa_{1}),\label{eq:2_PO_theta_2} \\
    \Dot{\kappa}_{2}&=-\epsilon(b\cos\theta+\kappa_{2}).\label{eq:2_PO_theta_3} 
\end{alignat}
\label{eq:2_PO_theta}
\end{subequations}
To determine the critical manifold, we set the (fast) layer equation Eq.~\eqref{eq:2_PO_theta_1} to zero:
\begin{equation*}
\begin{aligned}
    0&=\omega-\left(\kappa_1\sin(\theta+\alpha)+\kappa_2\sin(\theta-\alpha)\right)\\
    &=\omega-c_1\sin\theta-c_2\cos\theta,
\end{aligned}
\end{equation*}
with $c_1=\left(\kappa_1+\kappa_2\right) \cos\alpha$ and $c_2=\left(\kappa_1-\kappa_2\right) \sin\alpha$.
This yields the condition which describes the critical manifold:
\begin{equation}
    0=\omega-A(\kappa_1,\kappa_2)\sin(\theta+\gamma(\kappa_1,\kappa_2)),
    \label{eq:CM_cond}
\end{equation}
with $A=\sqrt{c_1^2+c_2^2}$ and $\gamma=\arg\left(c_1 +i c_2\right)$. 
For $\omega<A$ this yields two possible solutions: $\theta_1^*=\arcsin{\frac{\omega}{A}}-\gamma$ and $\theta_2^*=\pi-\arcsin{\frac{\omega}{A}}-\gamma$ with $\theta_1^*$ being stable and $\theta_2^*$ unstable. Here the stability is determined with respect to the fast dynamics \eqref{eq:2_PO_theta_1}.
Hence, the slow flow on the stable part of the critical manifold reads:
\begin{equation}
\begin{aligned}
    \Dot{\kappa}_1&=-\epsilon\left(-a\sin\left(\arcsin\frac{\omega}{A}-\gamma\right)+\kappa_1 \right),\\
    \Dot{\kappa}_2&=-\epsilon\left(b\cos\left(\arcsin\frac{\omega}{A}-\gamma\right)+\kappa_2 \right).
\end{aligned}
\end{equation}

\section{Averaging of the two phase oscillator\\ system}

For the case $\left|\omega\right|>A$, there is no equilibrium solution for Eq. \eqref{eq:CM_cond}. 
Instead, with fixed coupling weights, the layer equation \eqref{eq:2_PO_theta_1}  exhibits oscillations with frequency $\Omega=\sqrt{\omega^2-A^2}$, see \cite{FRA20}. 
The averaged flow of the coupling weights is obtained by averaging the adaptation functions over the period $T$ of the fast oscillations:
\begin{equation*}
\begin{aligned}
    {\kappa}'_i=\frac{1}{T}\int_0^{T}W_i(\theta(t))dt.
\end{aligned}
\end{equation*}
Here $\kappa_1$ and $\kappa_2$ are assumed to be constant, and $\theta(t)$ is the solution of \eqref{eq:2_PO_theta_1}, which depends implicitly on $\kappa_1$ and $\kappa_2$.
For the integral, we can use the layer equation \eqref{eq:2_PO_theta_1} to change the integration variable to $\theta$.
This yields the following expressions $W_1=-a\sin\theta-\kappa_1$ and $W_2=-b\cos\theta-\kappa_2$ and, hence:
\begin{subequations}    
\begin{alignat}{2}
        {\kappa}'_1&=-\left(\kappa_1-\Omega a\int_0^{2\pi}\frac{\sin\theta d\theta}{\omega-c_1\sin\theta-c_2\cos\theta}\right),\label{eq:int1}\\
        {\kappa}'_2&=-\left(\kappa_2+\Omega b\int_0^{2\pi}\frac{\cos\theta d\theta}{\omega-c_1\sin\theta-c_2\cos\theta}\right).\label{eq:int2}
\end{alignat}
\end{subequations}
The integrals are analytically solvable, and we show it with the example of Eq.~\eqref{eq:int1}. 
We rewrite the integral term in the following form:
\begin{equation}
\begin{aligned}[b]
&\int_{0}^{2\pi}\frac{\left(e^{i\theta}-e^{-i\theta}\right)d\theta}{2i\omega-c_{1}\left(e^{i\theta}-e^{-i\theta}\right)-ic_{2}\left(e^{i\theta}+e^{-i\theta}\right)}   \\
=&\int_{S^{1}}\frac{-zdz}{2\omega z+ic_{1}(z^{2}-1)-c_{2}(z^{2}+1)}-\int_{S^{1}}\frac{ydy}{2\omega y+ic_{1}(1-y^{2})-c_{2}(1+y^{2})}\label{eq:sin_int}
\end{aligned}
\end{equation}
with $z=e^{i\theta}$ and $y=e^{-i\theta}$. Both integrals can be solved using the residue theorem which we show for the integral containing $z$. The poles of the integrated function are:
\begin{equation*}
    \begin{aligned}
    z_{\pm} =	\frac{1}{(ic_{1}-c_{2})}\left(-\omega\pm\sqrt{\omega^{2}-c_{1}^{2}-c_{2}^{2}}\right).
        \end{aligned}
\end{equation*}
Since the residues must lie within the 1-circle, we compute their absolute value; recall that $A^2=c_1^2+c_2^2$:
\begin{equation*}
    \begin{aligned}
\left| z_{\pm}\right| = \frac{1}{A}\left|(-\omega\pm\sqrt{\omega^{2}-A^2})\right|.
    \end{aligned}
\end{equation*}
For $\omega>0$, 
the solution $z_+$ is located within the unit circle $|z_+|\le 1$.
We can therefore rewrite the first integral in Eq. \eqref{eq:sin_int} in the following way
\begin{equation}
    \begin{aligned}[b]
        & \int_{S^{1}}\frac{-zdz}{2\omega z+ic_{1}\left(z^{2}-1\right)-c_{2}\left(z^{2}+1\right)}
        =\int_{S^{1}}\frac{-zdz}{\left(ic_{1}-c_{2}\right)\left(z-z_{+}\right)\left(z-z_{-}\right)}\\
        &=\frac{-2\pi i}{\left(ic_{1}-c_{2}\right)}\frac{z_{+}}{z_{+}-z_{-}}
        =\frac{\pi i}{\left(ic_{1}-c_{2}\right)}\frac{\left(\omega-\sqrt{\omega^{2}-A^2}\right)}{\sqrt{\omega^{2}-A^2}}.
    \end{aligned}
\end{equation}
For the second integral in Eq. \eqref{eq:sin_int}, the same procedure yields
\begin{equation}
    \int_{S^{1}}\frac{-ydy}{2\omega y+ic_{1}\left(1-y^{2}\right)-c_{2}\left(1+y^{2}\right)}=\frac{\pi i}{\left(ic_{1}+c_{2}\right)}\frac{\left(\omega-\sqrt{\omega^{2}-A^2}\right)}{\sqrt{\omega^{2}-A^2}}.
\end{equation}
We can now write Eq. \eqref{eq:sin_int} as
\begin{equation*}
        \frac{\pi i}{\left(ic_{1}-c_{2}\right)}\frac{\left(\omega-\sqrt{\omega^{2}-A^2}\right)}{\sqrt{\omega^{2}-A^2}}+\frac{\pi i}{\left(ic_{1}+c_{2}\right)}\frac{\left(\omega-\sqrt{\omega^{2}-A^2}\right)}{\sqrt{\omega^{2}-A^2}}
        =\frac{2\pi c_{1}}{A^2}\frac{\left(\omega-\sqrt{\omega^{2}-A^2}\right)}{\sqrt{\omega^{2}-A^2}},
\end{equation*}
and, using $T=\frac{2\pi}{\sqrt{\omega^2-A^2}}$, the averaged flow for $\kappa_1$ reads
\begin{equation}\label{eq:averaged_flow_k1}
    {\kappa}'_1=\frac{c_1 a}{A^2}\left(\omega-\sqrt{\omega^{2}-A^2}\right)-\kappa_1 .
\end{equation}
The same procedure can be used for Eq.~\eqref{eq:int2} and yields
\begin{equation}\label{eq:averaged_flow_k2}
    {\kappa}'_2=\frac{-c_2 b}{A^2}\left(\omega-\sqrt{\omega^{2}-A^2}\right)-\kappa_2.
\end{equation}

\section{Bifurcation analysis of two coupled \\ phase oscillators}
\begin{figure}
    \centering
    \includegraphics[width=\textwidth]{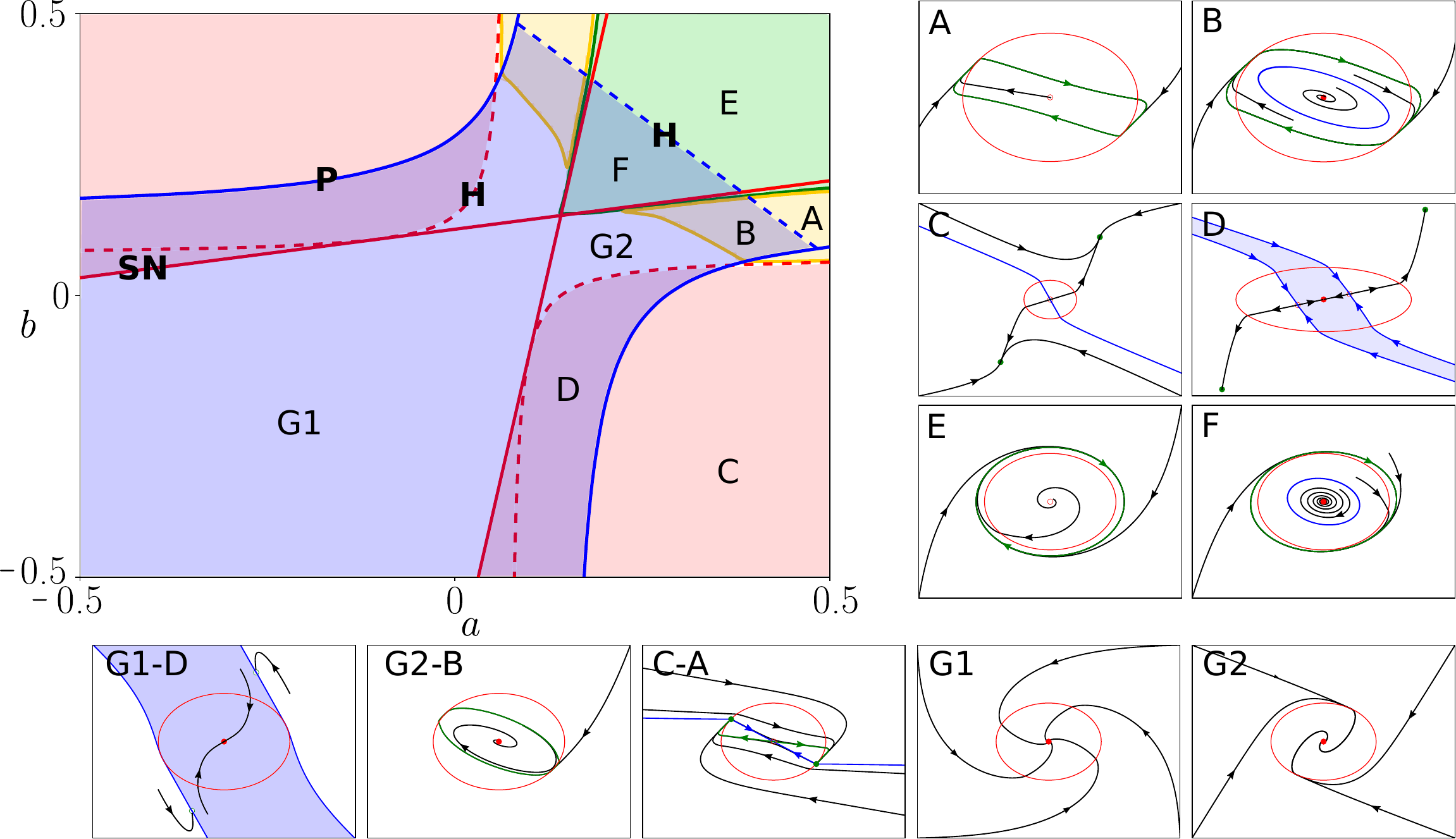}
    \caption{\textbf{Bifurcation diagram for the reduced system (10).} Bifurcation diagram in the parameter plane $(a,b)$ for $\omega=0.1$ and $\alpha=0.25\,\pi$. The regions G1 and G2 (blue) mark areas where the equilibrium $(\kappa_1,\kappa_2)=(0,0)$ is stable; C (red): a stable equilibrium on the critical manifold; A (orange): a stable recurrent synchronization limit cycle; E (green): a stable limit cycle on the critical manifold. 
    The other regions mark areas of bistability of the equilibria described above. 
    Also shown are analytically obtained bifurcation lines for the whole system (red) and the (0,0) equilibrium in the reduced system (blue), with SN denoting a Saddle-Node bifurcation, PF a Pitchfork bifurcation and H a Hopf bifurcation (dashed lines). 
    The boundaries for regions A, B, E and F are obtained numerically.
    The diagrams in the two columns on the right and the row in the bottom show exemplary trajectories for the reduced system in the $(\kappa_1,\kappa_2)$ plane for the regions denoted in the bifurcation diagram. Black lines mark stable trajectories, blue lines unstable trajectories and green lines limit cycle. The (0,0) equilibrium can either be stable (red and filled) or unstable (red and void). The blue areas denote the basins of attraction of the (0,0) equilibrium in the case of bistability. Green dots indicate stable fixed points on the critical manifold and the red line marks the boundary of the critical manifold.}
    \label{fig:bif_diag}
\end{figure}
\begin{figure}
    \centering
    \includegraphics[width=\textwidth]{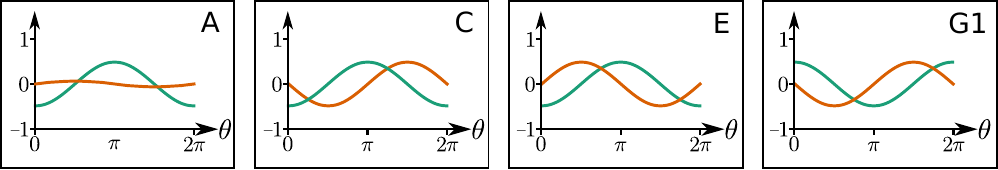}
    \caption{\textbf{Phase dependence of exemplary adaptation functions for different regions in the bifurcation diagram in Fig. \ref{fig:bif_diag}}. $\theta$-dependence of exemplary adaptation functions for the regions A ($a=0.07$, $b=-0.5$), C ($a=-0.5$, $b=-0.5$), E ($a=0.5$, $b=-0.5$) and G1 ($a=-0.5$, $b=0.5$) with the function influencing $\kappa_1$ shown in orange and the function acting on $\kappa_2$ shown in green.}
    \label{fig:plasticity_functions}
\end{figure}

Fig. 4a in the main text shows part of a bifurcation diagram for the two phase oscillator case (6)--(9). 
The complete bifurcation diagram can be seen in Fig.~\ref{fig:bif_diag}. 
Furthermore, exemplary trajectories of the reduced system in the $(\kappa_1, \kappa_2)$ plane are shown including the stable states for this region of the parameter space. 
Exemplary plasticity functions for the states A, C, E and G1 are shown in Fig. \ref{fig:plasticity_functions}

The bifurcation lines (red and blue), shown in the bifurcation diagram in Fig. \ref{fig:bif_diag}, result from the stability analysis of the fixed points in the original system (red lines) and of the $(\kappa_1, \kappa_2)=(0,0)$ fixed point in the averaged system (10). 
To compute the red bifurcation lines, we first compute the Jacobian $J$ of the equilibrium of the original system Eqs. \eqref{eq:2_PO_theta}
\begin{equation*}
   J= \left(\begin{matrix}-\kappa_{1}\cos\left(\alpha+\theta\right)-\kappa_{2}\cos\left(\alpha-\theta\right) & -\sin\left(\alpha+\theta\right) & \sin\left(\alpha-\theta\right)\\
a\epsilon\cos\theta & -\epsilon & 0\\
b\epsilon\sin\theta & 0 & -\epsilon
\end{matrix}\right),
\end{equation*}
and the corresponding characteristic equation
\begin{equation}\label{eq:3D_system_char_pol_t}
\begin{aligned}
    0=&(-\kappa_{1}\cos\left(\alpha+\theta\right)-\kappa_{2}\cos\left(\alpha-\theta\right)-\lambda)(\epsilon+\lambda)^{2} \\
    &+\sin\left(\alpha-\theta\right)b\epsilon\sin\theta(\epsilon+\lambda)-\sin\left(\alpha+\theta\right)a\epsilon\cos\theta(\epsilon+\lambda).
\end{aligned}
\end{equation}
For the bifurcation lines, the real part of an eigenvalue has to be zero, i.e., $\lambda=i\xi$:
\begin{equation}\label{eq:3D_system_char_pol}
\begin{aligned}
    0=&(-\kappa_{1}\cos\left(\alpha+\theta\right)-\kappa_{2}\cos\left(\alpha-\theta\right)-i\xi)(\epsilon+i\xi)^{2} \\
    &+\sin\left(\alpha-\theta\right)b\epsilon\sin\theta(\epsilon+i\xi)-\sin\left(\alpha+\theta\right)a\epsilon\cos\theta(\epsilon+i\xi).
\end{aligned}
\end{equation}
Setting $\xi=0$ allows us to determine the bifurcation lines for the real eigenvalues:
\begin{equation*}
\begin{aligned}[b]
    0=&	a\sin\theta\cos(\alpha+\theta)-b\cos\theta\cos(\alpha-\theta)+b\sin\theta\sin(\alpha-\theta)+a\cos\theta\sin(\alpha+\theta),\\
    0=&	a\sin(2\theta+\alpha)-b\cos(2\theta-\alpha).\\
\end{aligned}
\end{equation*}
For the case $\left(2\theta-\alpha\right)\neq0$ and $a\neq0$, this results in the following system:
\begin{equation*}
    \begin{aligned}
        a=&\frac{2\omega+b\cos\theta\sin(\theta-\alpha)}{\sin\theta\sin(\theta+\alpha)},\\
        b=&\frac{-2\omega\sin(2\theta+\alpha)}{\sin\theta\sin(\theta+\alpha)\cos(2\theta-\alpha)-\cos\theta\sin(\theta-\alpha)\sin(2\theta+\alpha)}.
    \end{aligned}
\end{equation*}
Here we have used the fixed point conditions from Eqs. \eqref{eq:2_PO_theta} to substitute $\kappa_1$, $\kappa_2$ and subsequently $a$.\\
For the case of $\cos\left(2\theta-\alpha\right)=0$ and $\alpha\neq\pi/4$, this yields two points:
\begin{equation*}
\begin{aligned}
    a=&	0,\\
    b_\pm=&	\frac{4\omega}{\sin\alpha\pm 1}.
\end{aligned}
\end{equation*}
For $\alpha=\pi/4$, $\sin(2\theta+\alpha)$ equals $\cos(2\theta-\alpha)$ which yields two lines in the case of $a\neq-b$: 
\begin{equation*}
    \begin{aligned}
        b_\pm=\frac{4\omega-a\left(\frac{1}{\sqrt{2}}\mp 1\right)}{\frac{1}{\sqrt{2}}\pm 1}.
    \end{aligned}
\end{equation*}
For $a=-b$, there exists one solution with $a=-2\sqrt{2}\omega$ and $b=2\sqrt{2}\omega$. This analysis yields the solid red lines in the bifurcation diagram of Fig. \ref{fig:bif_diag}.\\
For the case $\xi\neq 0$, Eq. \eqref{eq:3D_system_char_pol} yields a complex valued equation. The real part of the equation is
\begin{equation}\label{eq:3D_hopf_real}
\begin{aligned}
0=&\left(-a\sin\theta\cos\left(\alpha+\theta\right)+b\cos\theta\cos\left(\alpha-\theta\right)\right)\left(\epsilon^{2}-\xi^{2}\right)\\
&+2\epsilon\xi^{2}-\epsilon^{2}b\sin\left(\theta\right)\sin\left(\alpha-\theta\right)+\epsilon^{2}a\cos\left(\theta\right)\sin\left(\alpha+\theta\right).
\end{aligned}
\end{equation}
We can determine $\xi^2$ by using the equation for the imaginary part
\begin{equation}\label{eq:3D_hopf_imag}
    \begin{aligned}[b]
        \xi^{2}=&\epsilon\left(a\sin\theta\cos\left(\alpha+\theta\right)-b\cos\theta\cos\left(\alpha-\theta\right)+a\sin(2\theta+\alpha)-b\cos(2\theta-\alpha)\right)+\epsilon^{2}.
    \end{aligned}
\end{equation}
Plugging the fixed point condition for $\theta$ (Eq. \eqref{eq:2_PO_theta}) and Eq. \eqref{eq:3D_hopf_imag} into Eq. \eqref{eq:3D_hopf_real} yields a non-linear equation for $b$. This can be solved numerically for every $\theta$. The results are shown as dashed red lines in the bifurcation diagram in Fig. \ref{fig:bif_diag}.\\
The same procedure was done for the $(\kappa_1,\kappa_2)=(0,0)$ fixed point in the averaging regime (Eqs. \eqref{eq:averaged_flow_k1} and \eqref{eq:averaged_flow_k2}) resulting in the blue lines. For the case of $\xi=0$ (solid blue lines in Fig. \ref{fig:bif_diag}) this yields the following relation between $a$ and $b$:

\begin{equation*}
    b=\frac{-a\cos\alpha+2\omega}{\frac{-a}{\omega}\cos\alpha\sin\alpha+\sin\alpha}.
\end{equation*}

For the case of $\xi\neq 0$ the relation between $a$ and $b$ can also be given explicitly as long as the condition $\frac{ab\sin\alpha\cos\alpha}{2\omega^{2}}-\frac{a\cos\alpha+b\sin\alpha}{2\omega}+1\geq0$ holds. This results in:
\begin{equation*}
        b=-a\frac{\cos\alpha}{\sin\alpha}+\frac{4\omega}{\sin\alpha},
\end{equation*}
yielding the dashed blue lines in the bifurcation diagram of Fig. \ref{fig:bif_diag}.

\section{Period of the recurrent synchronization limit cycle}

\begin{figure}[H]
    \centering
    \includegraphics[width=0.6\textwidth]{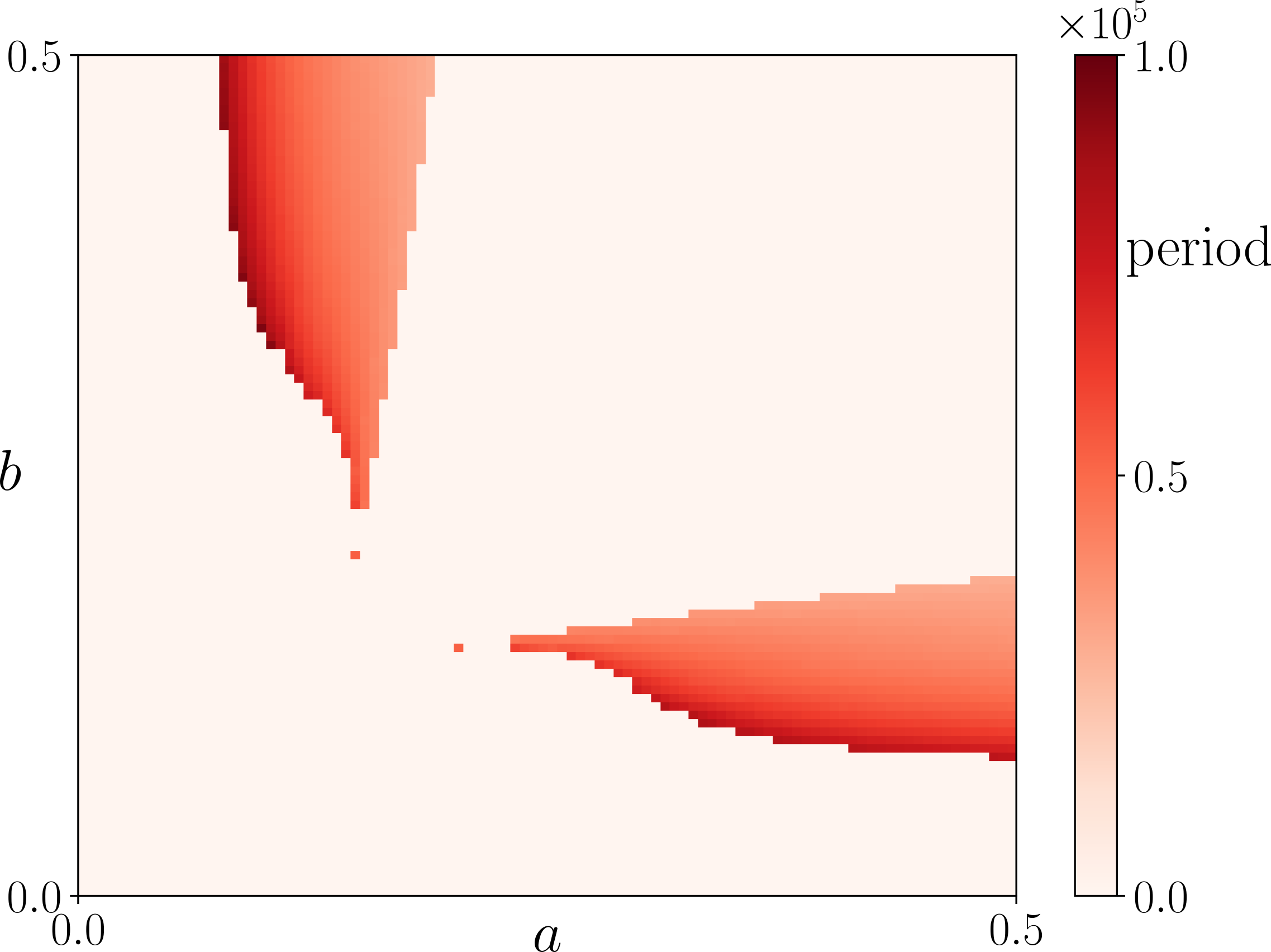}
    \caption{\textbf{Period of the recurrent synchronization limit cycle.} Periods of the recurrent synchronization limit cycle (color-coded) in the parameter-plane $(a,b)$ for $\omega=0.1$,  $\alpha=0.25\,\pi$, $\beta=-0.5\,\pi$ and $\epsilon=0.0001$. The light areas correspond to parameter combinations where no recurrent synchronization can be observed.}
    \label{fig:2PO_rs_period}
\end{figure}

To check, whether the recurrent synchronization limit cycle contains additional information about the bifurcation present in the system the periods are shown in Fig.~\ref{fig:2PO_rs_period}. Here the periods are color-coded in the parameter plane $(a.b)$, with the light areas signifying no recurrent synchronization. We can identify higher periods for lower values of $a$ and $b$ and vice versa. Since $a$ and $b$ denote the amplitude of the phase influence on the dynamics of the coupling weights, they therefore impact the speed with which trajectories move in the reduced phase space of the coupling weights $(\kappa_1,\kappa_2)$. This yields an inverse relationship between the values of $a$, $b$ and the period of the limit cycle.

\section{Influence of the time scale separation}

\begin{figure}[H]

    \centering
    \includegraphics[width=\textwidth]{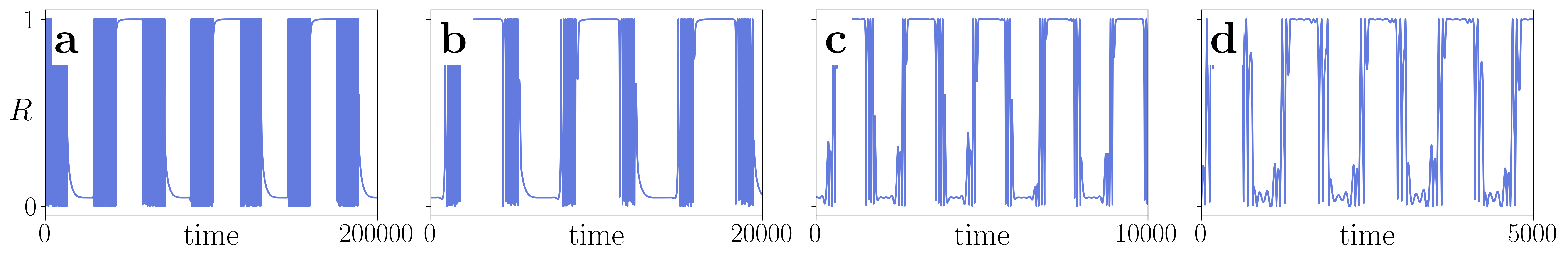}
    \caption{\textbf{Influence of the time scale separation.} Periods of recurrent synchronization for $\epsilon=0.0001$ (panel \textbf{a}), $\epsilon=0.001$ (panel \textbf{b}), $\epsilon=0.005$ (panel \textbf{c}) and $\epsilon=0.01$ (panel \textbf{d}). Remaining parameters: $a=0.5$, $b=0.1$, $\omega=0.1$, $\alpha=0.25\,\pi$, $\beta=-0.5\pi$. }
    \label{fig:2PO_diff_eps}
\end{figure}

To check the importance of the time scale separation (signified by the parameter $\epsilon$) for the occurrence of recurrent synchronization we show periods of recurrent synchronization for different $\epsilon$ in Fig.~\ref{fig:2PO_diff_eps}. We can see alternating episodes of phase locking and asynchrony for values from $\epsilon=0.0001$ (panel \textbf{a}) to $\epsilon=0.01$ (panel \textbf{d}) and inbetween (panels \textbf{b} and \textbf{c}). For higher values of $\epsilon$ recurrent synchronization was not observed, substantiating the importance of time scale separation. 

\section{Influence of asymmetry parameter $\beta$}

\begin{figure}[H]
    \centering
    \includegraphics[width=0.5\textwidth]{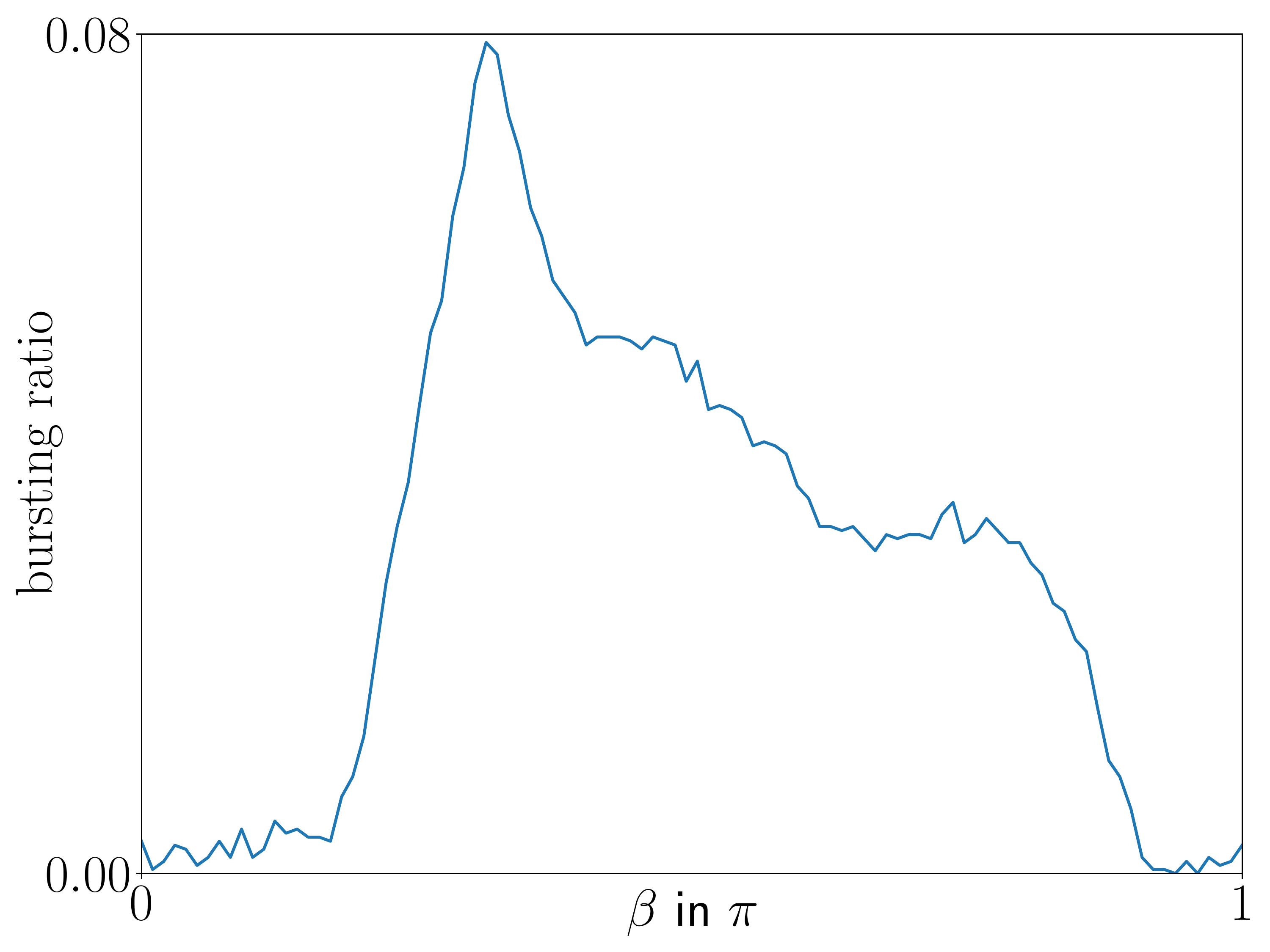}
    \caption{\textbf{Bursting ratio as a function of asymmetry $\beta$.} Ratio of stable recurrent synchronization to all states in the system of two phase oscillators (6)-(9) in the $(a, b)$-grid shown in Fig.~\ref{fig:bif_diag} for different values of $\beta$. Other parameters: $\epsilon=0.0001$ $\omega=0.1$, and $\alpha=0.25\,\pi$. }
    \label{fig:burst_ratio}
\end{figure}

To show that asymmetry is necessary for the emergence of recurrent synchronization, we compute the relative bursting ratio for different values of $\beta$ to determine how often recurrent synchronization occurs. This is done by producing the bifurcation diagrams shown in Fig. \ref{fig:bif_diag} and determining the ratio between the grid points with the recurrent synchronization (regions A and B) and the grid points of the whole parameter space. The result is shown in Fig.~\ref{fig:burst_ratio}.
We clearly observe that recurrent synchronization exists for a wide range of values of $\beta$ and is not constrained to the special value of $\beta=0.5\,\pi$. Furthermore, a certain amount of asymmetry in the adaptation functions seems to be necessary to obtain a stable recurrent synchronization. 

\section{Robustness of recurrent synchronization in networks of Hodgkin-Huxley neurons}

\begin{figure}
    \centering
    \includegraphics[width=\textwidth]{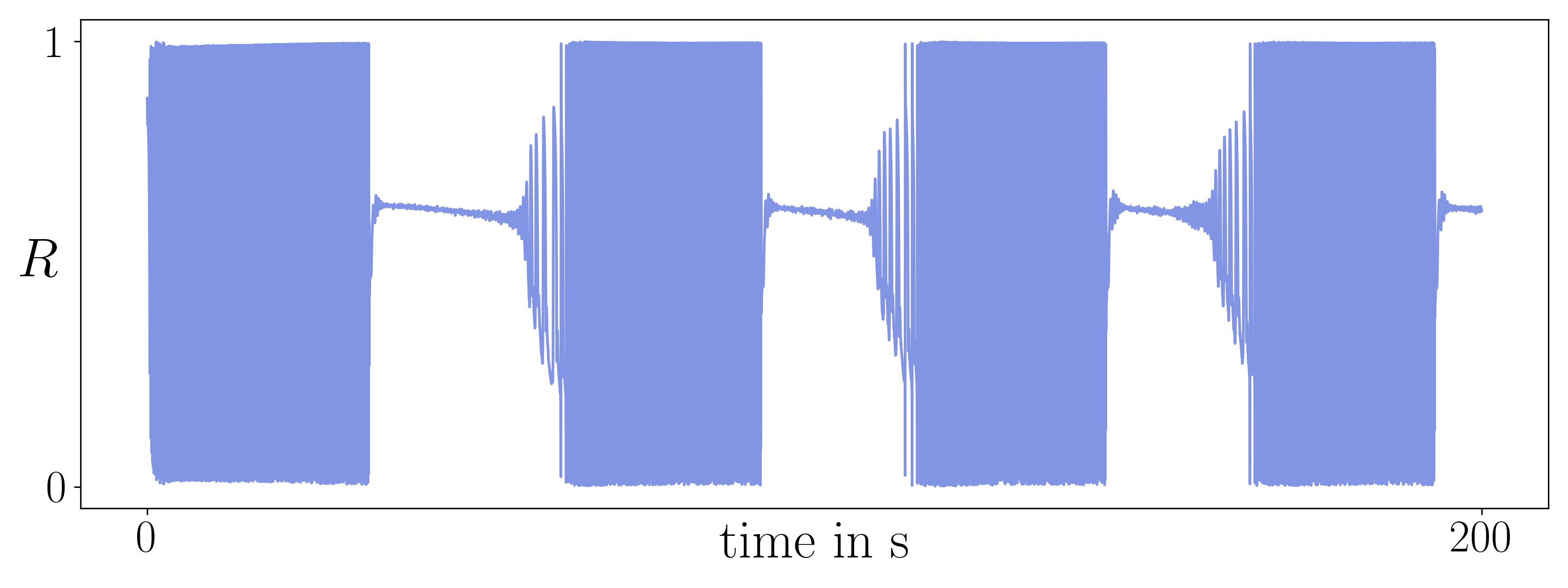}
    \caption{\textbf{Recurrent synchronization in a network of Hodgkin-Huxley neurons with an independent random input.} 
    Time evolution of the order parameter for 200 Hodgkin-Huxley neurons with noisy input from an $\alpha$-train defined in Eq.~\eqref{eq:alpha_train}. 
    Initial conditions for the coupling weights and voltages are randomly chosen from the intervals $[0, 0.5]$ and $[-70\, \textnormal{mV}, 20\, \textnormal{mV}]$, respectively. 
    The input currents are $5\,\mu \mathrm{A}$ and $13\,\mu \mathrm{A}$ for populations 1 and 2 and the noise amplitude for the $\alpha$-train is $I=0.01$. The remaining parameters are as described in section 4.4.}
    \label{fig:HHN_alpha_train}
\end{figure}

\begin{figure}
    \centering
    \includegraphics[width=\textwidth]{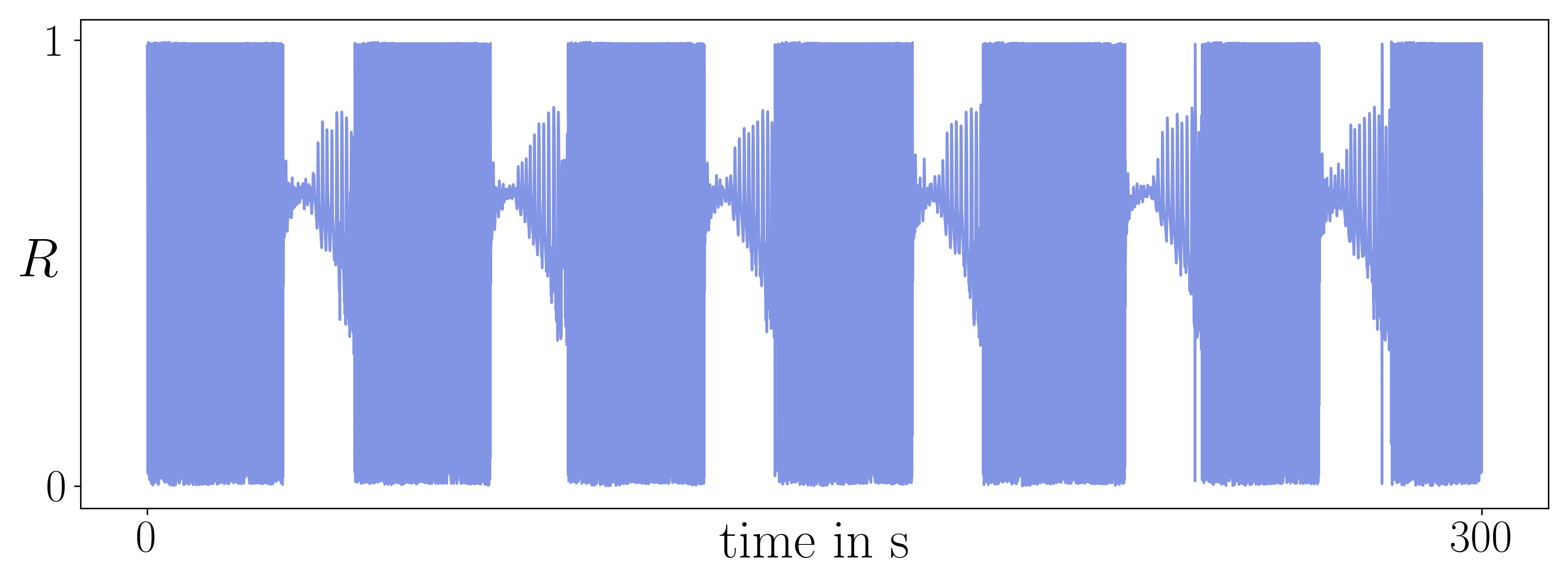}
    \caption{\textbf{Recurrent synchronization in a network of Hodgkin-Huxley neurons with heterogeneous constant currents.} 
    Time evolution of the order parameter for 200 Hodgkin-Huxley neurons after transient. 
    The input currents for population 1 and 2 are chosen from the intervals $[4.5\, \mathrm{\mu A}, 5.5\, \mathrm{\mu A}]$ and $[12.5\, \mathrm{\mu A}, 13.5\, \mathrm{\mu A}]$, respectively. 
    The initial conditions for the coupling weights and the starting voltage were randomly chosen from the intervals $[0, 0.5]$ and $[-70\, \textnormal{mV}, 20\, \textnormal{mV}]$.}
    \label{fig:HHN_delta_omega_05}
\end{figure}

To investigate the robustness of the recurrent synchronization in two populations of Hodgkin-Huxley neurons (Sec. 2.2), we simulate a network of $N=200$ neurons where every neuron receives an independent input in the voltage dynamics. 
We choose an $\alpha$-train input of the form \cite{POP13}
\begin{equation}\label{eq:alpha_train}
I_{i}^{\text {input }}(t)=I\left(V_{r}-V_{i}(t)\right) \sum_{\tau_{i, k}<t} \alpha\left(t-\tau_{i, k}\right) e^{-\alpha\left(t-\tau_{i, k}\right)},
\end{equation}
modeling random post-synaptic potentials arriving at the $i$th neuron with the time intervals between two potentials being distributed as $\mathcal{N}(14\,\textnormal{ms},4\,\textnormal{ms})$. The time evolution of the order parameter for $I=0.01$ can be seen in Fig. \ref{fig:HHN_alpha_train}. Recurrent synchronization is still present, but the oscillations in the phase-locked periods have larger amplitude than in the noiseless case shown in Fig. 2b of the main text. 
Furthermore, the results shown in Fig. 2c of the main text are also present for increased heterogeneity in the input current each neuron receives, which is shown in Fig.~\ref{fig:HHN_delta_omega_05}. Here we show the order parameter of the network for the last $300\,$s of a $600\,$s long simulation for expanded intervals around the input currents of $5\,\mathrm{\mu A}$ for population 1 and $13\,\mathrm{\mu A}$ for population 2 from $0.02\,\mathrm{\mu A}$ to $1\,\mathrm{\mu A}$. Recurrent synchronization is still present in this more heterogeneous system and for the longer time interval.

\begin{figure}[H]
    \centering
    \includegraphics[width=\textwidth]{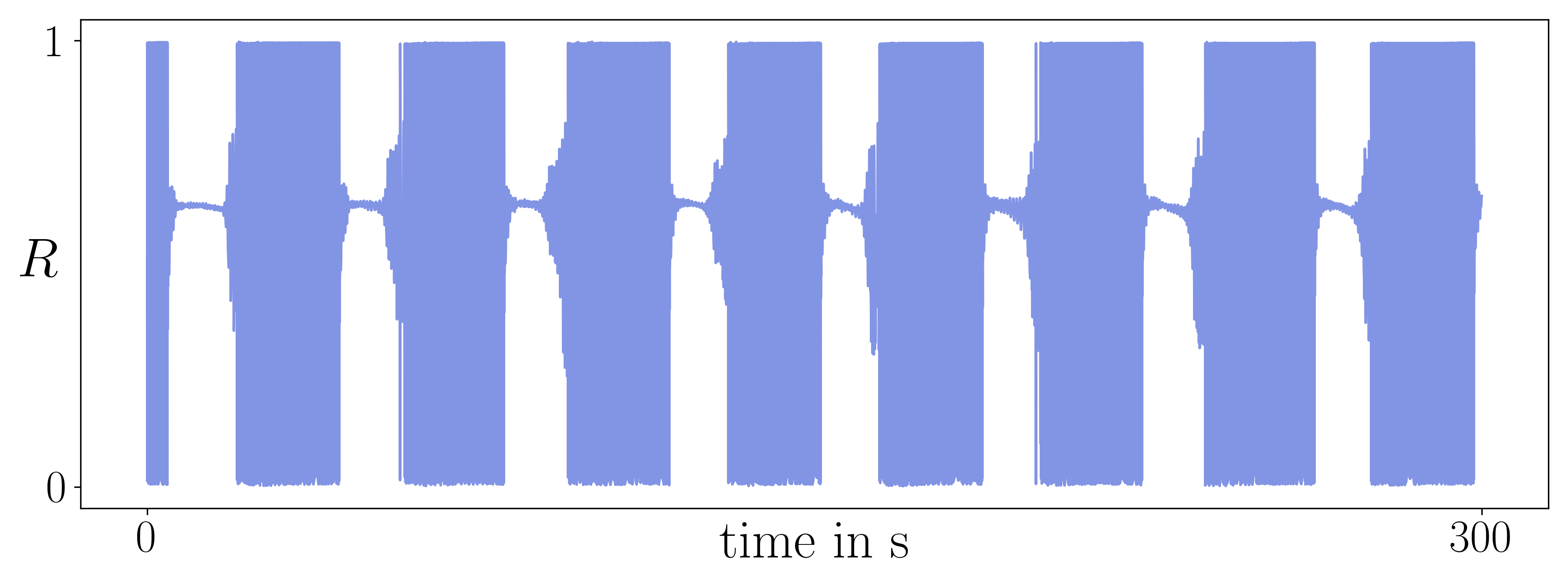}
    \caption{\textbf{Recurrent synchronization in a network of Hodgkin-Huxley neurons with heterogeneous coupling weights update functions.} 
    Time evolution of the order parameter for 200 Hodgkin-Huxley neurons with $\tau_1$ from $W_1$ distributed on the interval $[0.24,0.26]$ and $\gamma$ from $W_2$ distributed on the interval $[0.95,1.05]$. 
    Initial conditions for the coupling weights and voltages are randomly chosen from the intervals $[0, 0.5]$ and $[-70\, \textnormal{mV}, 20\, \textnormal{mV}]$, respectively. 
    The input currents for population 1 and 2 are chosen from the intervals $[4.99\, \mathrm{\mu A}, 5.01\, \mathrm{\mu A}]$ and $[12.99\, \mathrm{\mu A}, 13.01\, \mathrm{\mu A}]$, respectively. The remaining parameters are as described in section 4.4.}
    \label{fig:HHN_network_dist_adap_functions}
\end{figure}

Fig. 3e shows the occurrence recurrent synchronization for different update functions of the coupling weights in the 2 neuron case. In Fig.~\ref{fig:HHN_network_dist_adap_functions} the order parameter for a network 200 neurons with distributed update functions is shown. Recurrent synchronization is again present even for case of distributed input currents, update functions and initial conditions.

\section{The case of equal plasticity functions for two Hodgkin-Huxley neurons}

\begin{figure}[H]
    \centering
    \includegraphics[width=\textwidth]{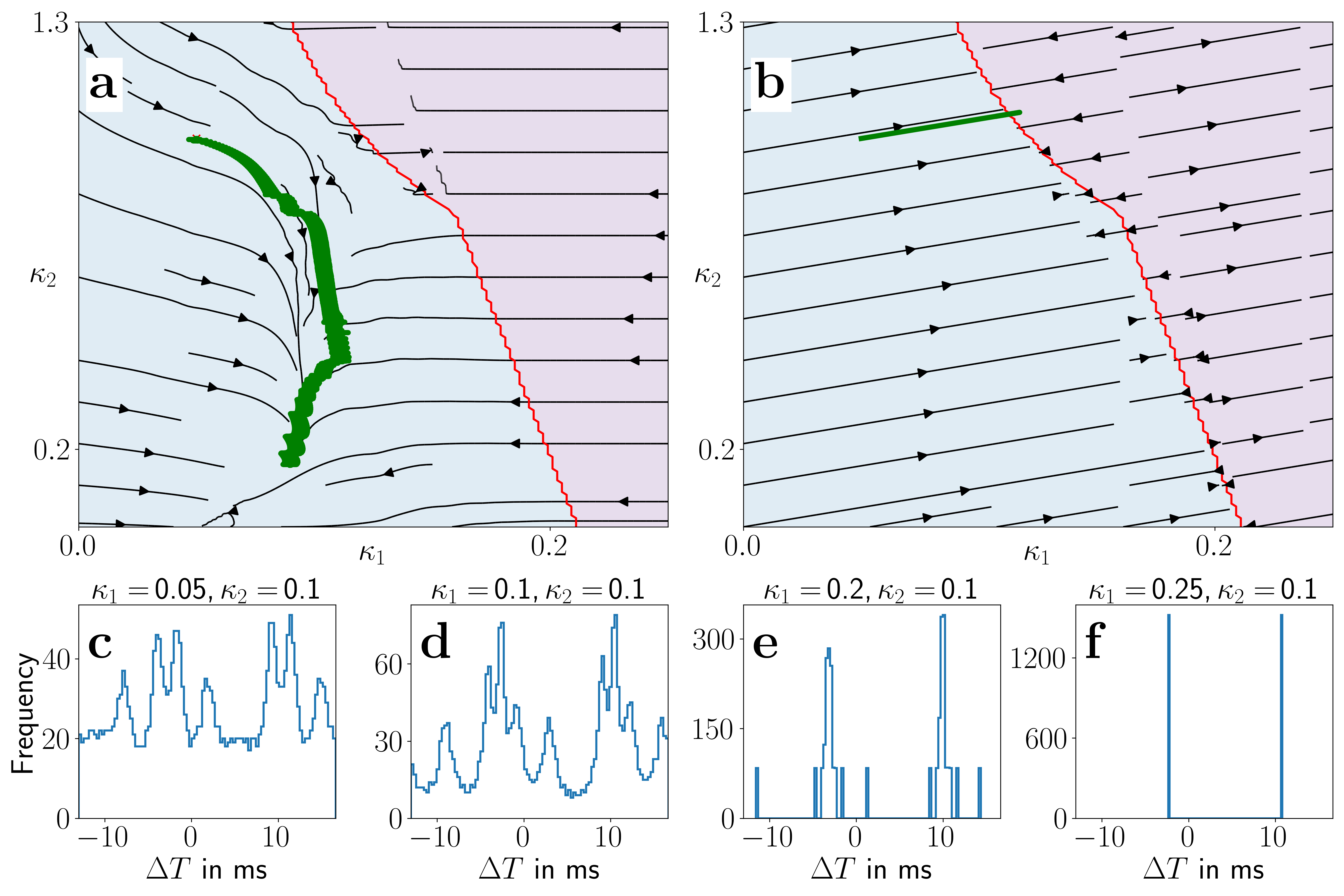}
    \caption{\textbf{Two Hodgkin-Huxley neurons with identical update functions do not display recurrent synchronization.} 
    \textbf{a} and \textbf{b}: reduced flow for two Hodgkin-Huxley neurons with identical update functions of the coupling weights; \textbf{a} displays the case of both coupling weights being updated by $W_1$ given by Eq. (12) of the main text; \textbf{b} shows the case of identical update functions $W_2$ given by Eq. (13) of the main text. 
    The green curve is the projection of the trajectory of the whole system (starting in $(\kappa_1,\kappa_2)=(0.05,1)$ in both cases) onto the $(\kappa_1,\kappa_2)$-plane and the red line marks the boundary between the synchronous (light pink) and asynchronous (gray blue) regions. 
    \textbf{c - f}: distributions of the timing difference between the spikes of neurons 1 and 2 (relative to the spiking times of neuron 1) for fixed coupling weights $\kappa_2=0.1$ and $\kappa_1=0.05$, $0.1$, $0.2$, and $0.25$, which corresponds to the change from  asynchronous to synchronous regime.}
    \label{fig:2HHN_sym_W}
\end{figure}

Figure~\ref{fig:2HHN_sym_W} illustrates how the system of two Hodgkin-Huxley neurons evolves when the update functions for $\kappa_1$ and $\kappa_2$ are identical.
For the considered parameter values, we observe no recurrent synchronization. 

Shown is the trajectory of the whole system (in green) projected onto the $(\kappa_1,\kappa_2)$-plane with both update functions being equal to $W_1$ defined in  Eq.~(12) (Fig.~\ref{fig:2HHN_sym_W}a) and to $W_2$ defined in Eq.~(13) of the main text (Fig.~\ref{fig:2HHN_sym_W}b). These trajectories are in very good agreement with the reduced flow based on the averaging approach described in section 4.4 (black lines). Both the trajectory and the flow show no presence of a periodic orbit traversing the boundary between the synchronous and asynchronous spiking regimes, which would be the way recurrent synchronization appears in this representation. 

Figure~\ref{fig:2HHN_sym_W} also displays exemplary distributions of the spike timing differences: Figs.~\ref{fig:2HHN_sym_W}c and  \ref{fig:2HHN_sym_W}d  for  asynchronous regimes, Fig.~\ref{fig:2HHN_sym_W}e near the boundary,  and Fig.~\ref{fig:2HHN_sym_W}f for the synchronous regime, with the two peaks consistent with phase-locked spiking. These figures illustrate the impact the coupling weights have on the relative spiking times.

\end{document}